\documentclass[twocolumn]{aastex62}

\usepackage{xcolor}
\usepackage{textcomp}
\definecolor{green}{rgb}{0.31, 0.78, 0.47}
\makeatletter
\def\@makefnmark{\hbox{(\normalfont\@thefnmark)}}
\makeatother

\begin{document}

\title{Near-infrared Spectral Characterization of Solar-Type Stars in the Northern Hemisphere}

\author{Collin D. Lewin}
\affiliation{Steward Observatory, University of Arizona, 933 North Cherry Avenue, Tucson, AZ 85721, USA; \href{mailto:clewin@mit.edu}{clewin@mit.edu}}

\author{Ellen S. Howell}
\affiliation{Lunar and Planetary Laboratory, University of Arizona, 1629 East University Boulevard, Tucson, AZ 85721, USA}

\author{Ronald J. Vervack, Jr.}
\affiliation{Johns Hopkins Applied Physics Laboratory, 11100 Johns Hopkins Road, Laurel, MD 20723, USA}

\author{Yanga R. Fern\'andez}
\affiliation{University of Central Florida, 4000 Central Florida Boulevard, Orlando, FL 32816, USA}

\author{Christopher Magri}
\affiliation{University of Maine at Farmington, 224 Main Street, Farmington, ME 04938, USA}

\author{Sean E. Marshall}
\affiliation{Arecibo Observatory/UCF, PR-625, Arecibo 00612, PR, USA}

\author{Jenna L. Crowell}
\affiliation{NAVWARSYSCOM, San Diego, CA, USA}

\author{Mary L. Hinkle}
\affiliation{University of Central Florida, 4000 Central Florida Boulevard, Orlando, FL 32816, USA}

\begin{abstract}
Although solar-analog stars have been studied extensively over the past few decades, most of these studies have focused on visible wavelengths, especially those identifying solar-analog stars to be used as calibration tools for observations. As a result, there is a dearth of well-characterized solar analogs for observations in the near-infrared, a wavelength range important for studying solar system objects. We present 184 stars selected based on solar-like spectral type and V\textminus J and V\textminus K colors whose spectra we have observed in the 0.8-4.2 micron range for calibrating our asteroid observations. Each star has been classified into one of three ranks based on spectral resemblance to vetted solar analogs. Of our set of 184 stars, we report 145 as reliable solar-analog stars, 21 as solar analogs usable after spectral corrections with low-order polynomial fitting, and 18 as unsuitable for use as calibration standards owing to spectral shape, variability, or features at low to medium resolution. We conclude that all but 5 of our candidates are reliable solar analogs in the longer wavelength range from 2.5 to 4.2 microns. The average colors of the stars classified as reliable or usable solar analogs are V\textminus J=1.148, V\textminus H=1.418, and V\textminus K=1.491, with the entire set being distributed fairly uniformly in R.A. across the sky between \textminus27 and +67 degrees in decl.
\end{abstract}

\section{Introduction} \label{sec:intro}
Many observations of solar system objects rely on solar-analog stars in order to analyze the reflected sunlight measured at the telescope. Spectral observations designed to characterize asteroids, in particular, are mostly carried out in visible or near-infrared spectral regions and can provide measurements of size and composition.\\
\indent Over the past 10 years, we have been observing near-Earth asteroids using the NASA Infrared Telescope Facility (IRTF) to better understand their surface composition and thermal properties. Near-Earth asteroids near 1 AU have significant thermal flux in the 3-5 micron spectral region. With the IRTF/SpeX instrument \citep{2003PASP..115..362R},  we measured spectra between 0.8-2.5 microns and 1.9-4.2 microns using two different modes of the instrument. After the 2014 IRTF upgrade, these wavelength ranges were extended to 0.7-2.5 microns and 1.7-5.3 microns respectively. The combined wavelength coverage encompasses the transition from purely reflected light to a thermally dominated regime. Analysis of the spectra requires careful calibration to separate the effects of viewing geometry, changing heliocentric distance, and rotation of the asteroid from the true variability across the asteroid's surface, which may be present. For this work, we use relative reflectance spectra, defined as the ratio of the asteroid spectrum to that of a solar-type star, normalized appropriately. Although working in absolute flux would have advantages, the significant disadvantage of needing photometric conditions and additional absolute flux calibration would limit our program to too few objects. Instead, we use relative reflectance spectra and model the entire spectral range in a self-consistent way, which has been very successful as shown in \cite{2018Icar..303..220H}, \cite{2018Icar..303..203M}, and \cite{2017Icar..292...22M}. \\
\indent Near-Earth asteroids can appear anywhere in the sky during times of close approach, so we have needed solar-analog stars with a wide sky distribution. On each observing night, we have chosen stars near the asteroid position at the time of observation based on their catalog colors, trying to match those of the Sun. In addition, we also observed at least one well-characterized primary solar-analog star---chosen from the published literature---in order to check the nearby star. On any given night, we compare all the stars to the primary solar analog, and compare each asteroid with each star (further details follow in Section \ref{sec:data}). Over the course of this project, we have observed 184 solar-type stars selected based on spectral type and V\textminus J and V\textminus K colors closely resembling those of the Sun. For these observations, stars with $6 < V < 10$ are suitable, with $7 < V < 9$ magnitudes best to reach adequate signal-to-noise at all wavelengths of interest and to prevent saturation in the shorter-wavelength range (low-resolution mode of SpeX). \\
\indent We aimed to expand our list of well-characterized solar stars as well as produce a catalog of good near-infrared solar-analog stars all over the northern sky. We chose an initial set of 8 primary solar analogs for which we have ample observations, with most being included in the lists of solar-analog stars presented in \cite{1992AJ....104..340L}, \cite{1985AJ.....90..896C}, or \cite{1980A&A....91..221H} and the rest being selected for our observations based on their solar-like colors from the Two Micron All-Sky Survey (2MASS), Hipparcos, and Tycho catalogs \citep{1997A&A...323L..49P, 2000A&A...355L..27H, 2003yCat.2246....0C}. Using our classification system (see Section \ref{subsec:scheme}), we have analyzed the spectra of the 184 stars we have observed, with the goal of reevaluating our initial list of 8 primary solar analogs. Here, we present a list of 17 well-characterized, primary solar analogs which we confirm are spectrally consistent over time and lack nonsolar spectral features at low to medium resolution. We also include a table of all 184 stars and their suitability as calibration stars, with spectral plots of all stars in the 0.8-4 micron region included as supplemental plots.

\section{Observations and data reduction} \label{sec:data}
\subsection{Near-infrared Star Observations}
As part of our observational program, we measured the spectra of many solar-type stars in order to correct for telluric absorption and the contribution of the solar spectrum in our collected asteroid spectra. All of our near-infrared star spectra were collected using SpeX \citep{2003PASP..115..362R} at the NASA IRTF on 133 nights spanning May 2008 to October 2018, in both prism (0.8-2.5 microns) and LXD1.9 (2.0-4.2 microns) modes. Spectra collected after August 2014 using post-upgrade SpeX cover an extended wavelength range of 0.7-5.3 microns between both the prism and LXD modes. We used the 0.8-arcsecond slit, which was usually comparable to or wider than the seeing at K-band (typically 0.5-1.0 arcseconds). In a few cases, we used the 1.5-arcsecond slit when the seeing exceeded an arcsecond. The spectral resolution is $\frac{ \lambda}{\Delta\lambda}=200$ in prism mode, and 2500 in LXD mode. The effective spectral resolution with the 0.8-arcsecond slit is 0.021 microns, and we sample the spectrum at a uniform 0.015 microns in the prism range. Similarly, in LXD mode the effective resolution is 0.0023 microns and sampled at 0.001 microns. During the observations, we guided using the light reflected from the slit jaws with the 1.0-2.5 micron imaging camera, with an autoguider keeping the object centered in the slit during spectral integrations. We averaged several pairs of observations in each set, typically 4-8 pairs of exposures in each mode, nodded along the slit to collect A and B spectra. Pre-upgrade, we used exposures of 1-10s in LXD1.9 mode to get good sky subtraction for the calibration stars. Post-upgrade, our exposures are limited to 3s and two coadds to avoid the sky saturating at wavelengths longer than 4 microns. Internal arc lamps were used for wavelength calibration, along with sky lines at the longer wavelengths. Spectral flat fields were taken with internal lamps, as described in \cite{2003PASP..115..362R}, at each telescope position, within about 1 hour of R.A. in case of flexure.
\begin{figure*}[htb!]
\epsscale{0.65}
\plotone{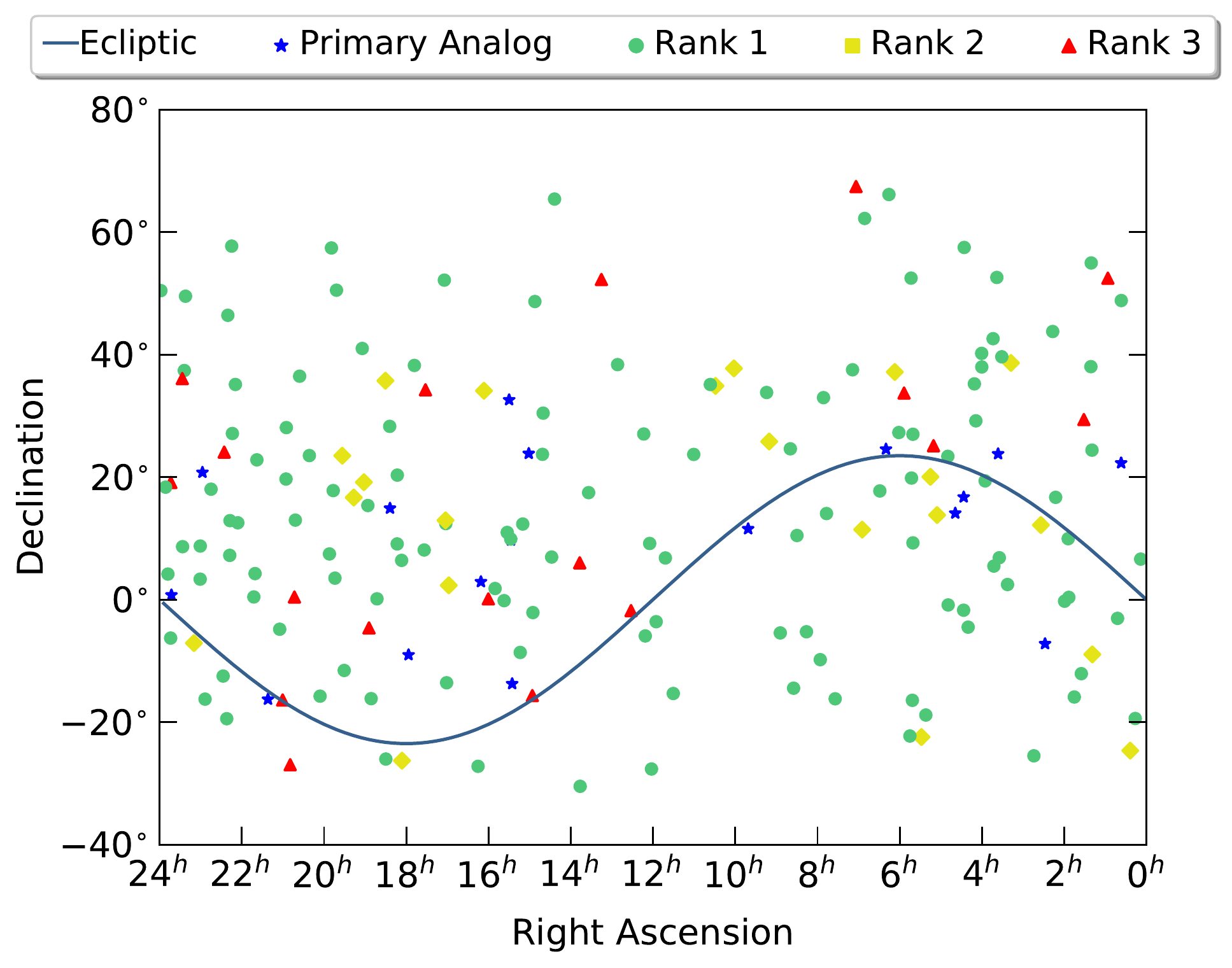}
\caption{Distribution of our solar-analog candidates across the sky plotted in equatorial coordinates with R.A. in hours and decl. in degrees, with the ecliptic superimposed in blue. Our set of stars spans the sky from \textminus27 to +67 degrees in decl. We have a hole in our distribution from 6 to 12 hours in R.A. and above 40 degrees in decl. for which we have no stars. This hole is further discussed in Section \ref{sec:discussion}. Each of the three rankings assigned to the stars with our classification scheme (see Section \ref{subsec:scheme}) is plotted in a different color reflecting the results shown in Table \ref{tab:thetable}. }
\label{fig:dist_sky}
\end{figure*}
\subsection{Data Reduction and Atmospheric Correction} \label{subsec:reduction}
A majority of our data reduction, including spectral extraction, consisted of processing our SpeX data with the Spextool software described by \cite{2004PASP..116..362C}. In both prism and LXD modes, the spatial profile perpendicular to the slit is generated along the profile, and the signal is corrected column by column and extracted along the spectral trace. The spectra are extracted from both nodded slit positions and averaged to make the final spectrum.

\indent We applied Bus's method for correcting telluric water vapor in the prism spectra to fit the telluric contribution by modeling the atmosphere with the known altitude and zenith angle at the time of observation. The details of this method are provided by \cite{2004Icar..172..408R}. A similar method was implemented to correct for telluric features in the LXD data and is detailed by \cite{2007Icar..187..464V}. Each star spectrum was matched to a modeled atmosphere with respect to both wavelength and the depth of telluric features \citep{1992nstc.rept.....L}. The spread of data points in these features was then iteratively minimized. The best-fit column depth of atmospheric water was determined for the 1.4-micron and 1.9-micron absorption bands in the prism data, with the average of these two values used to correct for the weaker 0.92-micron water feature. The 3.0-micron absorption band (2.8-3.7 microns) was used to fit the water column depth for the LXD data. Throughout these iterative processes, data points with values outside of 2$\sigma$ from surrounding points were flagged as bad and ignored in subsequent iterations. Both the prism and LXD spectrum of each candidate star were divided by that of a well-characterized, primary solar-analog star and often several additional candidate stars for our analysis and classification detailed below in Section \ref{sec:analysis}. All data points in the comparison spectra were normalized to unity at 1.65 microns for prism spectra and 2.35 microns for LXD spectra. The V\textminus J, V\textminus H, and V\textminus K colors---with the J, H, K filter passbands centered at 1.25, 1.65, and 2.20 microns, respectively---of both stars provided by the Hipparcos and 2MASS catalogs were converted from magnitudes to flux units, divided, and plotted with propagated errors in the comparison spectra for comparing the stars' colors. For stars bright enough to have likely saturated in the 2MASS fields, which is indicated by quality codes of C or worse, we utilize colors from VizieR catalog II/225 \citep{1993cio..book.....G (ADS)}. 

\section{Candidate Star Classification} \label{sec:analysis}
\subsection{Candidate Selection}
\begin{figure*}[ht!]
\epsscale{0.55}
\plotone{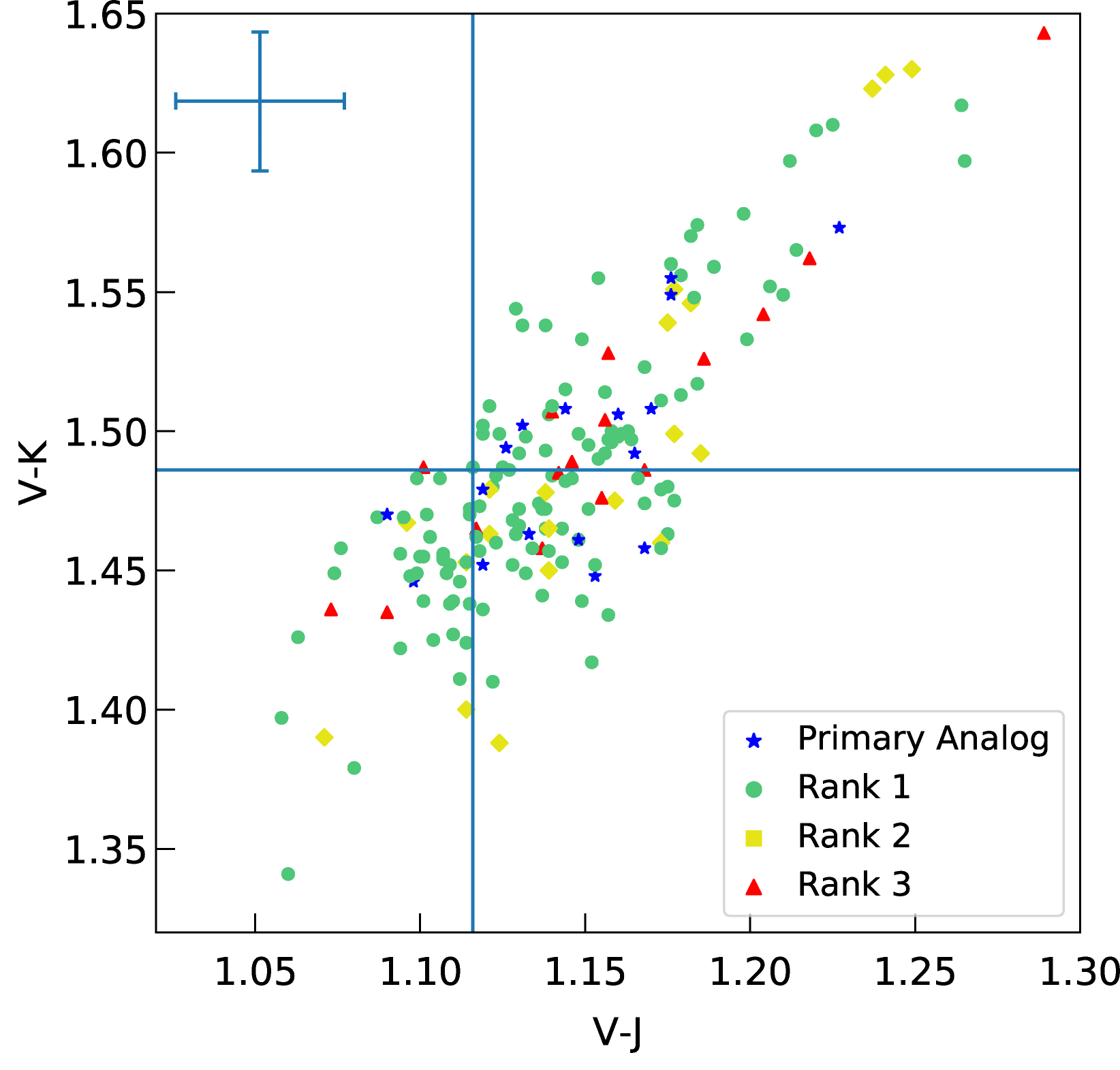}
\caption{V\textminus J and V\textminus K spectral colors of our set of solar analog candidates using photometry reported by the 2MASS (for J and K) and either Hipparcos or Tycho catalogs (for V). Assumed solar colors of V\textminus J=1.116 and V\textminus K=1.486 are shown by the blue vertical and horizontal lines, respectively. Each of the three rankings assigned to the stars with our classification scheme (see Section \ref{subsec:scheme}) is plotted in a different color reflecting the results shown in Table \ref{tab:thetable}. Seven stars (4 rank-1 stars, 1 rank-2 star, and 2 rank-3 stars) fall outside the axes limits, which are constrained for clarity. The typical uncertainties for both V\textminus J and V\textminus K is 0.025 magnitudes, as shown in the upper-left corner.}
\label{fig:dist_colors}
\end{figure*}

When observing many solar system objects, a solar-analog star is utilized to correct or remove the Sun's spectrum from the observed spectrum of reflected sunlight. As a result, such an analog must be readily observed through the same telescopic instrument. In our case of observing near-Earth asteroids, we need stars that are close to the target in the sky on the nights of observation. In the near-infrared, particularly in the 3-5 micron spectral range, the atmosphere changes rapidly on a time scale of minutes. The water vapor content can also change spatially, so the best calibration star is located near the target object in the sky so it can be observed frequently through the same atmospheric path. \\
\indent Near-Earth asteroids can appear in any part of the sky and move rapidly from night to night, motivating a need for a list of solar-type stars spread all over the northern sky and suitable for our observations on the NASA IRTF. Figure \ref{fig:dist_sky} shows the distribution across the sky of the stars we have used in our observations, which spans a decl. range of \textminus27 to +67 degrees. For each target on a given night, we search for stars with colors near that of the Sun and within an angular distance of 5 degrees of the asteroid at the time of observation. In a few cases, if no suitable stars are found, the distance is extended to 8-10 degrees. We also include a primary solar analog to compare the selected calibration star with a known solar-analog star. By ``primary solar analog" we mean a star which has been previously determined in the visible to be a good spectral match to the Sun and, after our observations, has been added to our list of primary solar analogs presented in Table \ref{tab:newset}. To avoid confusion between terminology, a ``solar twin" has previously been defined by \cite{1996A&ARv...7..243C} as a star with physical parameters (mass, chemical composition, metallicity, etc.) similar to, if not identical to, the Sun. We use this term in our paper for describing Hyades 106, which we find to be our most reliable primary solar analog. For the purposes of this paper, and for matching the spectral slopes in the near-infrared, primarily 1-2.5 microns, these other characteristics are less important. \\
\indent We have observed our set of primary solar analogs over many nights, allowing us to confirm their reliability, consistency, and lack of nonsolar spectral features, as further discussed in Section \ref{subsec:analogs}. We selected these stars to calibrate asteroid spectra, which do not usually have narrow spectral features at low to medium resolution. We prefer G0-G5 stars to avoid spectral features, but have found that the stellar spectra are not always as predicted (see Section \ref{sec:discussion}). Stars that showed features were not used as calibrators, so that spurious spectral features were not introduced into the asteroid spectra, and these stars are labeled ``rank 3" according to our classification scheme detailed below in Section \ref{subsec:scheme}. \\
\indent Our calibration stars were also selected for having V\textminus J and V\textminus K colors closely resembling those of the Sun, for which we assume solar colors of V\textminus J = 1.116, V\textminus  H = 1.426, and V\textminus K = 1.486 \citep{1985AJ.....90..896C}. We prioritize V\textminus K resemblance, typically selecting stars to match V\textminus J within $\pm0.08$ and V\textminus K to within $\pm0.06$. Figure \ref{fig:dist_colors} shows the V\textminus J, V\textminus K color distribution of our stars. Although a $6 < V < 10$ star is usable for our asteroid spectra calibration, $7 < V < 9$ is preferred for an adequate signal-to-noise ratio in a short period of time at long wavelengths, while still avoiding detector saturation at short wavelengths. The typical uncertainties for both V\textminus J and V\textminus K colors is 0.025 magnitudes, shown on the plot. Individual uncertainties for each star are given in Table \ref{tab:thetable}.

\subsection{Classification Scheme} \label{subsec:scheme} 
Our classification scheme for assessing a candidate's resemblance to the Sun consists in ranking the star's comparative spectra and using these results to assign a final ranking of the star. The classification system for both the individual spectra and, subsequently, the star itself are the same, with three ranks based on $\Delta$, the magnitude of the deviation of the relative slope from unity:
	
\begin{itemize}
  \item[] \fcolorbox{black}{green}{\rule{0pt}{3pt}\rule{3pt}{0pt}} \underline{Rank 1:} star is a reliable solar analog; no corrections needed ($\Delta < 10\%$ for prism spectra, $\Delta < 20\%$ for LXD spectra)
  \item[] \fcolorbox{black}{yellow}{\rule{0pt}{3pt}\rule{3pt}{0pt}}  \underline{Rank 2:} star requires polynomial-fitting corrections to be a usable solar analog ($10\% \leq \Delta < 20\%$ for prism spectra) 
  \item[] \fcolorbox{black}{red}{\rule{0pt}{3pt}\rule{3pt}{0pt}}   \underline{Rank 3:} star is effectively nonsolar in that it is not correctable by polynomial fitting owing to variability, nonsolar features, or spectral shape ($\Delta \geq 20\%$ for both prism and LXD spectra)
\end{itemize}

\begin{figure}[t]
\includegraphics[width=\linewidth]{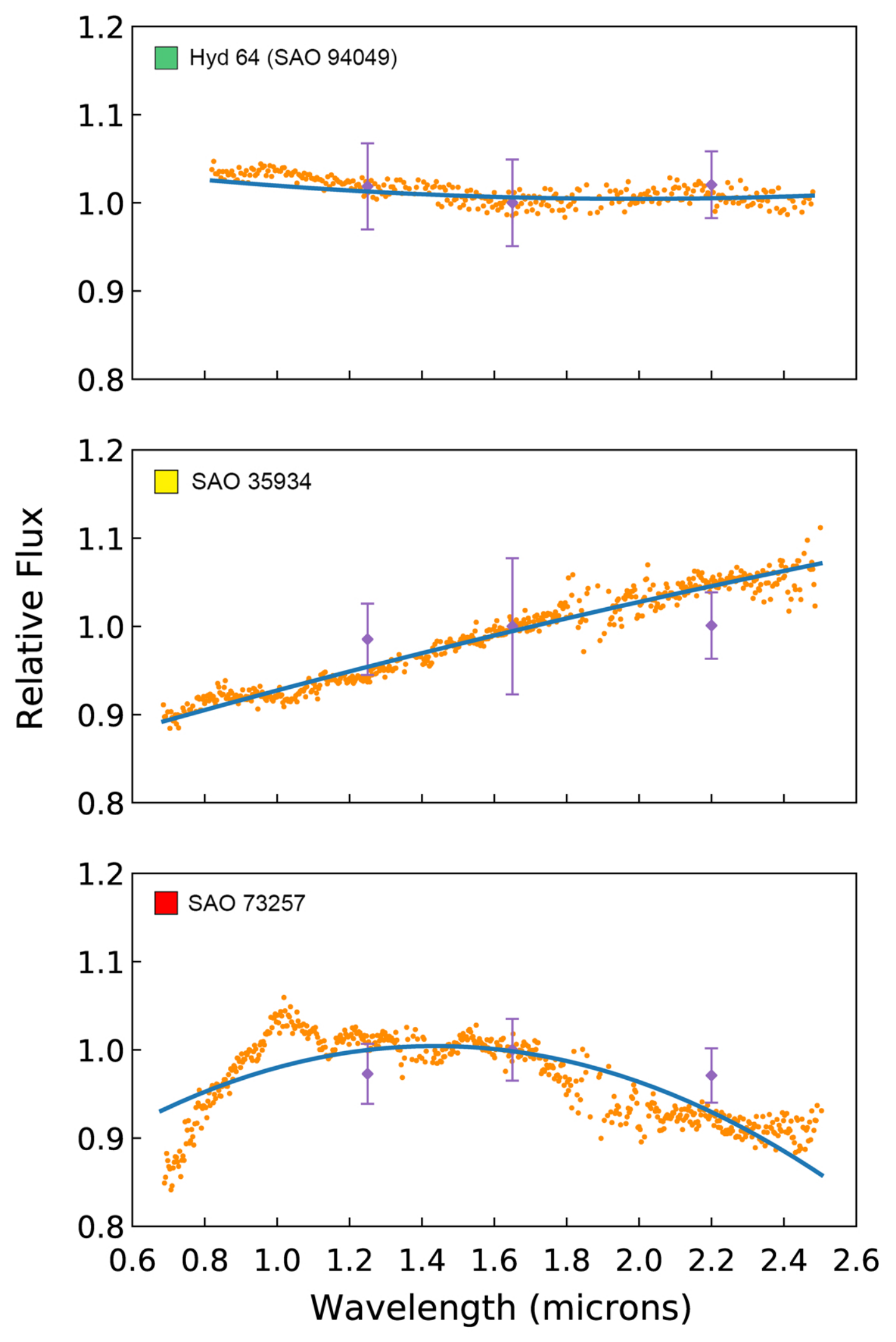}
\caption{Examples of comparison prism spectra comparing Hyades 64, SAO 35934, and SAO 73257 with a star from our list of primary solar analogs, each with a second-order polynomial fit shown in blue. Relative V\textminus J, V\textminus H, and V\textminus K colors are plotted with error bars determined from the reported uncertainty in the catalog colors. Each spectrum corresponds to one of the three rankings of our classification scheme outlined in Section \ref{subsec:scheme}, indicated by the color of the square matching that used in Table \ref{tab:thetable} (top: rank 1, middle: rank 2, bottom: rank 3). Explanations for each of the example classifications are also detailed in Section \ref{subsec:scheme}.}
\label{fig:explots}
\end{figure}

\begin{figure}[t]
\includegraphics[width=\linewidth]{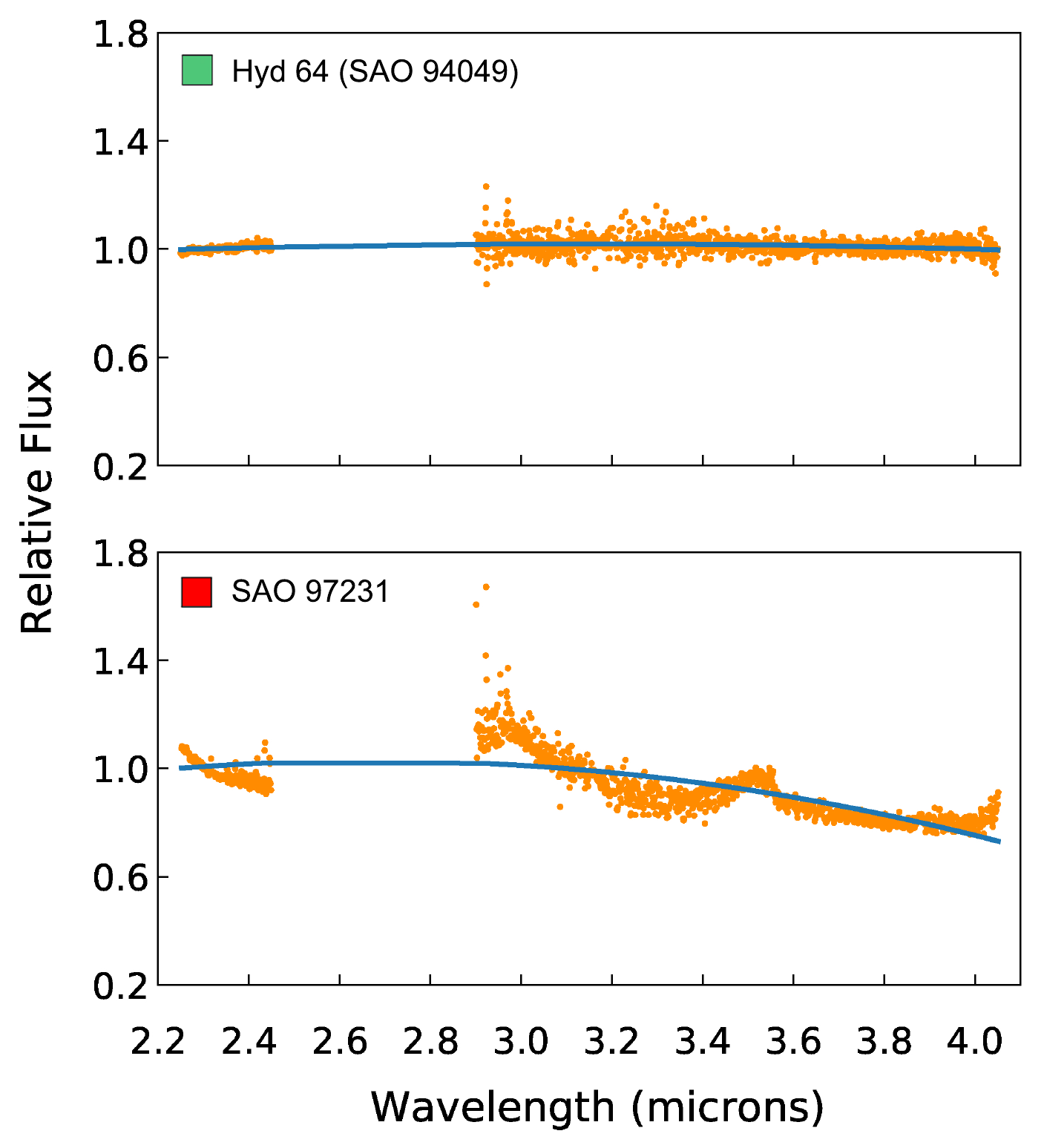}
\caption{Examples of comparison LXD spectra comparing Hyades 64 and SAO 97231 with a star from our list of primary solar analogs, each with a second-order polynomial fit shown in blue. Each spectrum corresponds to one of the two rankings for LXD spectra, indicated by the color of the square (top: green/rank 1, bottom: red/rank 3). The water feature from 2.45 to 2.90 microns has been omitted and is not used in the fitting.}
\label{fig:LXD_explots}
\end{figure}

We assess how similar the spectra of two stars are by determining the amount their comparison spectrum deviates from a flat line at 1.0, the result of dividing two identical spectra. Each comparison spectrum was first fit with a second-order polynomial fit, with the maximum deviation of the fit from a flat line at 1.0 being our statistic for quantifying the deviation from a flat line. The maximum percentage that the polynomial fit of the comparison spectrum deviates from a flat line at 1.0 is denoted above as $\Delta$. \\
\indent The threshold for determining if a spectrum requires a polynomial-fitting correction to be used is based on whether any slope in the spectrum causes the second-order polynomial fit to fall outside of 10\% from a flat line at 1.0 for the comparison prism spectra and 20\% for comparison LXD spectra. Prism spectra outside the 10\% threshold but within 20\% are assigned rank 2 if correctable with a low-order (3-4) polynomial fit or rank 3 if the spectral shape cannot be fit with a low-order polynomial. Stars showing significant variability on any time scale or nonsolar, low-to-medium resolution features are also assigned rank 3. Stars whose prism spectra are within the 10\% threshold and LXD spectra within 20\% and lack features and variability are ranked 1 and are considered to be reliable solar analogs. \\
\indent Choosing the threshold for maximum $\Delta$ for rank-1 spectra is arbitrary, but the $10\%$ threshold for prism and $20\%$ threshold for LXD are about twice the repeatability of measurements of the spectrum of the same star over several nights. An advantage of spectral observations is that absolute calibration is not needed, and thus usable data can be obtained on nights with cirrus or nonphotometric conditions. However, cirrus does reduce the near-infrared sky stability, and can lead to small spectral slope changes and incomplete  correction of telluric features from atmospheric lines being broader and sometimes saturated. The difference in this criterion for maximum deviation between rank-1 prism ($<10\%$) and LXD ($<20\%$) spectra is a consequence of the relationship between the signal-to-noise of the spectra in these two modes. The asteroid spectra at LXD wavelengths greater than about 3.5 microns is dominated by thermal emission, and so is relatively brighter by a factor of 10-100 than at 2 microns, despite having larger fractional uncertainties.\\
\indent By our criteria, all but 5 of our candidate solar-analog stars are rank 1 in the LXD range, so the primary focus of our classifications is on the prism wavelength range. Our final rank for each star is based on the prism spectra. We do not have LXD observations for 12 of our stars and have several stars for which our only LXD observations have too low signal-to-noise, resulting from star faintness and/or observing conditions, to adequately characterize the star at these wavelengths. In this case, we include a note in Table \ref{tab:thetable} to identify such stars. \\
\indent Classifications were first assigned to the prism and LXD spectra comparing the candidate star on each observational night to well-characterized, primary solar analogs. If all the comparisons of the candidate star's spectra to primary solar analogs are ranked 1, the candidate is assigned rank 1. However, if several of these spectra are rank 2, we classify the star using the average $\Delta$ calculated from those for the individual comparative spectra. Examples of a rank 1, rank 2, and rank 3 prism spectrum are provided in Figure \ref{fig:explots} for clarity. The rank 1 spectrum (top) is almost entirely flat, with the entire polynomial fit falling within $10\%$ from a flat line. The rank 2 spectrum (middle) has a clear positive slope, resulting in the left end of the polynomial fit falling just outside of $10\%$ but within $20\%$ from a flat line at 1.0. Although the bottom spectrum falls within the 20\% levels, there are distinct spectral features from 1.0 to 1.6 microns resulting in a spectral shape that deviates significantly from the polynomial fit; thus, we classify it as rank 3. A similar example for a rank 1 and rank 3 LXD spectrum (we remind the reader that there is no rank 2 for LXD spectra) is shown in Figure \ref{fig:LXD_explots}. The rank 1 spectrum (top) is flat, with the second-order polynomial fit falling within 20\%. The rank 3 spectrum (bottom) clearly exhibits prominent features centered at 2.95 and 3.5 microns. 

\subsection{Set of Primary Solar Analogs}\label{subsec:analogs}
\begin{figure*}
\includegraphics[width=\linewidth]{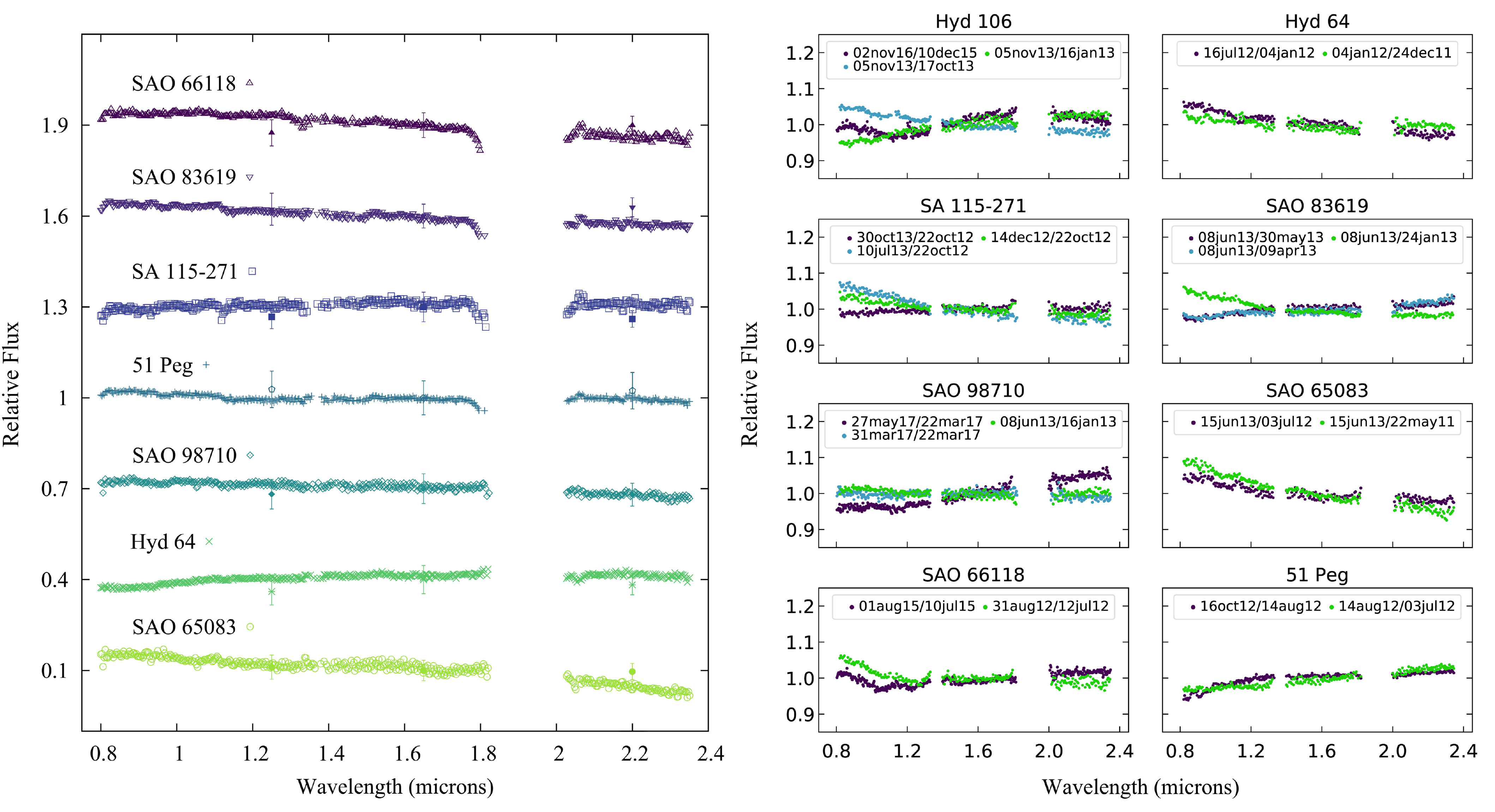}
\caption{(left) Comparative spectra from dividing each star in our original set of 8 primary solar analogs by a representative Hyades 106 spectrum, normalized to unity and shifted for clarity. As described in Section \ref{sec:analysis}, the results of dividing the V\textminus J, V\textminus H, and V\textminus K colors of each star (in relative flux units) by those of Hyades 106 is plotted with error bars. (right) Comparative spectra from dividing two spectra of the same star collected on different dates. Data leftwards of 0.8 microns has been removed due to instrumental sensitivity drops from 0.7-0.8 microns. The telluric water feature centered at 1.9 microns has been removed from all spectra and the 1.35-micron water feature from spectra shown on the right. The results and limitations of these comparisons are discussed in Section \ref{subsec:analogs}.}
\label{fig:starcomp}
\end{figure*}
\indent In order to assess the viability of each candidate star as a solar analog, we must first construct a set of reliable solar analogs to use in the evaluations. To that end, we assembled an initial set of 8 primary solar analogs, with the majority chosen from the lists of solar-analog stars presented by \cite{1992AJ....104..340L}, \cite{1985AJ.....90..896C}, or \cite{1980A&A....91..221H} and the remaining stars being selected for their solar-like colors from the 2MASS, Hipparcos, and Tycho catalogs. We have ample observations of this set of 8 stars, with the number of observations ranging from 6 to 33 nights and an average of 17 nights for the entire set. \\
\indent To determine whether or not our initial primary analogs are truly viable as solar analogs, we carried out two tests. First, we divided each primary analog's spectra by the spectra of the other seven to ensure that the stars resemble one another, as solar analogs should, and that they lack nonsolar features. Second, for a given star, we compared spectra from all the nights on which it was observed to verify that the star's spectrum is consistent over time. \\
\indent An example of the first comparison test is shown in Figure \ref{fig:starcomp}, where spectra of the seven other initial primary solar analogs have been divided by the spectrum of Hyades 106 (SAO 94049). After determining that Hyades 106 closely resembles Hyades 64 (SAO 93936), a long-known solar analog at visible wavelengths, and is slightly brighter and thus better fit for our asteroid calibrations, we have observed Hyades 106 copiously. Owing to this large number of observations of Hyades 106, we have chosen to designate it as the solar twin in our set (the best solar analog), which we used as the basis for comparison of the other seven stars in our initial set when assessing their reliability as solar analogs. \\
\indent For each star in the initial set of 8 primary solar analogs, a representative spectrum was chosen for the purposes of comparing to the other seven stars. This representative spectrum was selected as being the most typical spectrum for the star from all the nights on which it was observed. We remind the reader that our process for ranking candidate stars includes a comparison to more than one representative spectrum; instead, we compare all observations of a candidate to those of at least one of our primary solar analogs. We find that all of the spectra comparing each star to Hyades 106 are flat enough to be classified as rank 1 in our classification system (spectrum within $\pm 10\%$ of a flat line), except for the spectrum comparing SAO 65083 to Hyades 106. The red end of the comparative SAO 65083 spectrum deviates by just over 10$\%$, resulting in a rank 2 classification of this star. 51 Peg has been long used as a solar analog in the visible, but its 2MASS catalog colors differ by more than two standard deviations from those of Hyades 106.  However, 51 Peg was most likely saturated in the 2MASS fields as shown by the JHK quality codes of D, D, and C, respectively. We instead use the pre-2MASS IR photometry presented in VizieR catalog II/225, resulting in 51 Peg's colors being consistent with those of Hyades 106, as shown in Figure \ref{fig:starcomp}. We therefore find 51 Peg to be a rank-1 star and, thus, it stays in our list of primary solar analogs.
\\
\indent For the second test of checking for time variability, the representative spectrum for each star was compared to all the other spectra of that star collected throughout our observations. Similar to how we selected a characteristic spectrum of each primary solar analog in our set, the spectra used to compare a star to itself over time were chosen for being representative of the star's other spectra we collected within a month of the observation. As shown in the results of this spectral comparison over time in Figure \ref{fig:starcomp} (right), spectral variations are visible across a timescale of several months in a majority of the solar analogs. However, all of these variations are within $\pm10\%$ from being a straight line, meeting our criterion for assigning a comparison spectrum rank 1; a majority of these changes are within $5\%$.

\begin{figure}[t]
\includegraphics[width=\linewidth]{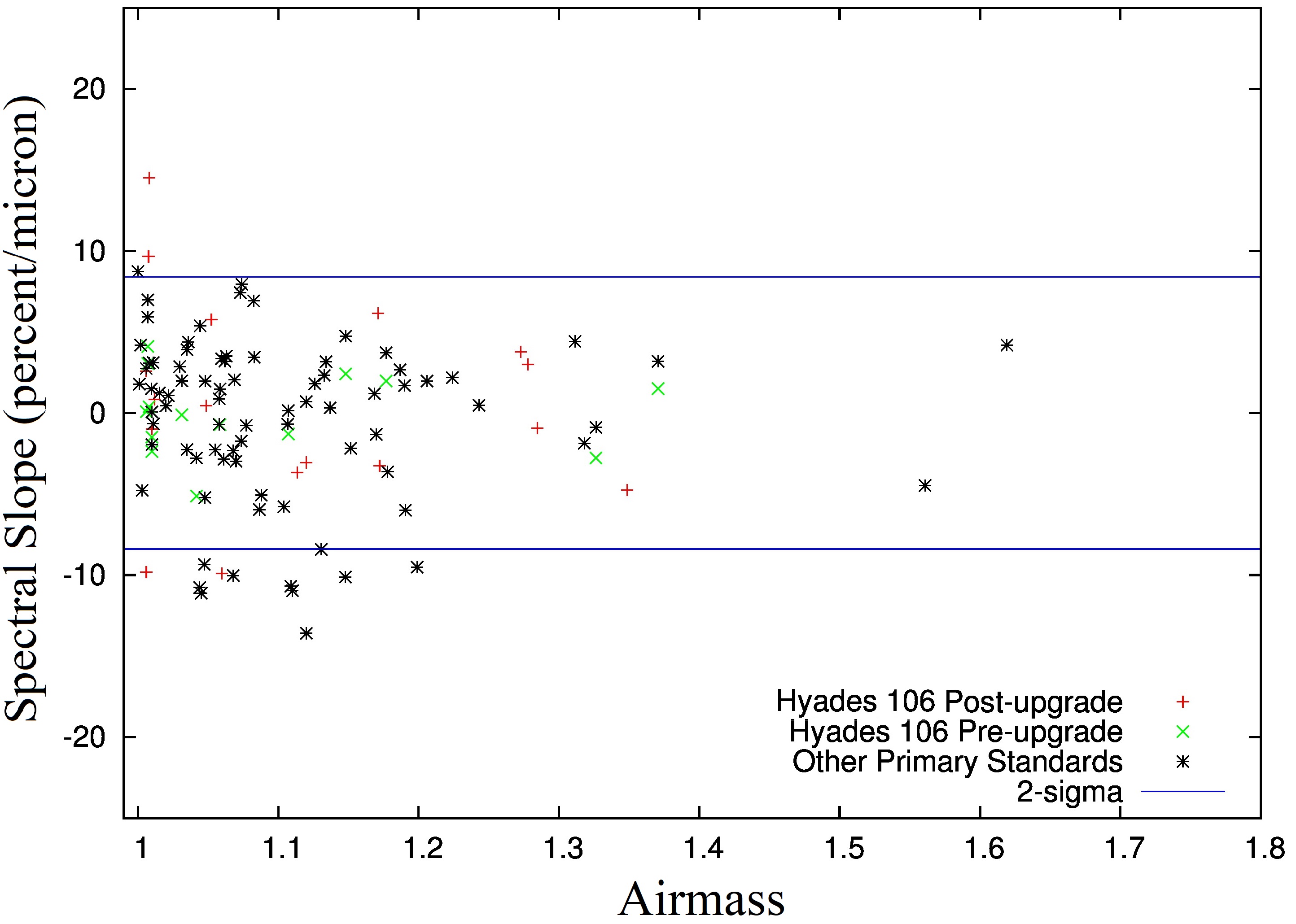}
\caption{Spectral slopes from 0.85 to 2.45 microns as a function of airmass for observations of Hyades 106 (red and green) and other primary standard stars (black) selected from Table \ref{tab:newset}. No particular trend arises with observing conditions, though most spectral slope variation occurs at low airmass rather than at high airmass.}
\label{fig:airmass}
\end{figure}

\begin{figure}[t]
\includegraphics[width = \linewidth]{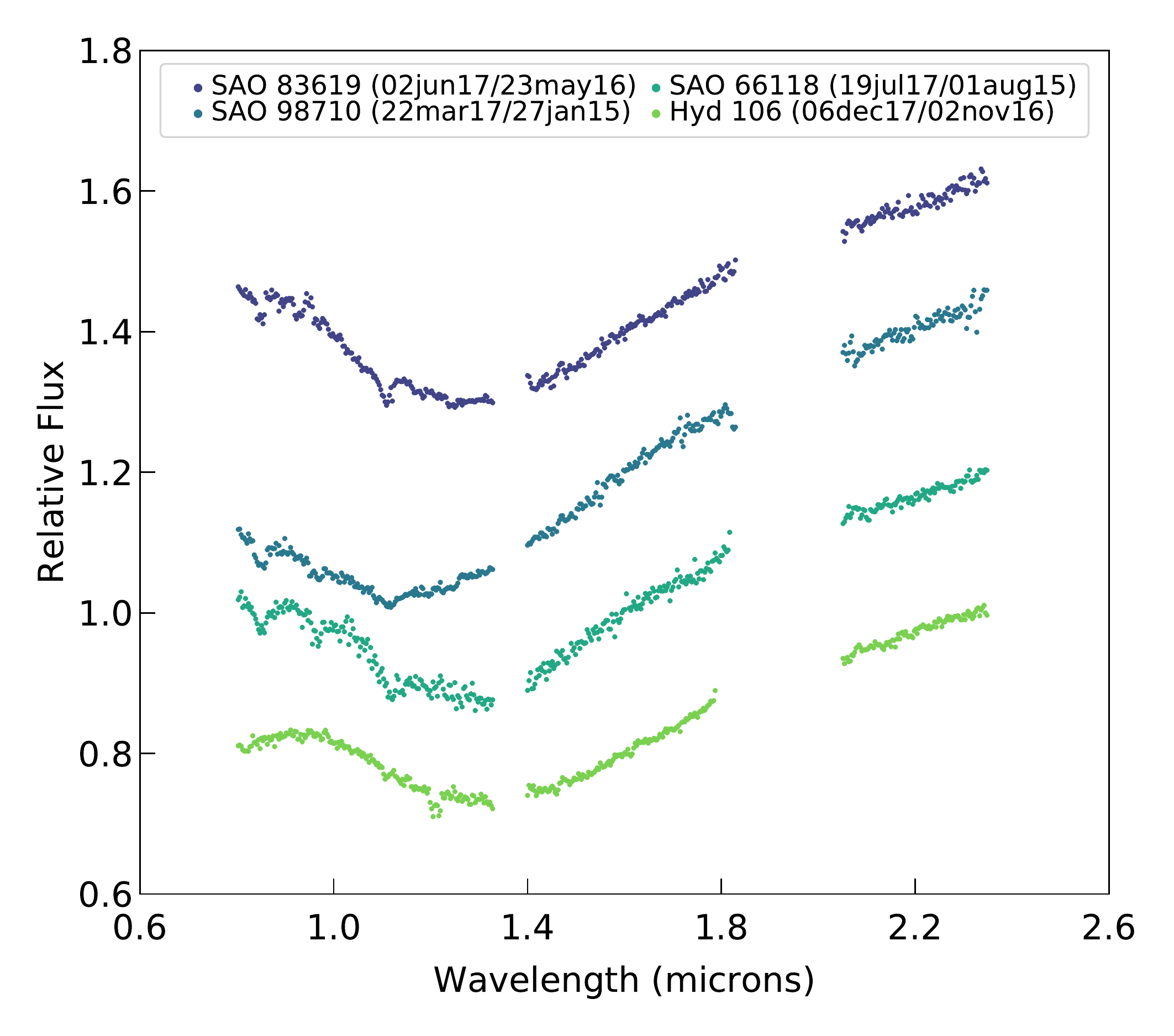}
\caption{Four examples of the anomalous pattern that often results when comparing two spectra collected more than a year apart following the 2014 IRTF SpeX upgrade. Each spectrum, normalized to unity and shifted for clarity, compares spectra of the same star collected on different nights (listed in brackets). We do not fully understand the cause of this effect but do not think it indicates any change in the stars themselves (see further discussion in Section \ref{subsec:analogs}).}
\label{fig:anomalies}
\end{figure}

\begin{deluxetable*}{cccccccc}[]
\tablecaption{Final Set of 17 Primary Solar Analogs, with V Photometry from the Hipparcos and Tycho Catalogs. \label{tab:newset}}
\tablewidth{55pt}
\tablehead{\colhead{Star} & \colhead{SAO Number} & \colhead{Alt. Name} & \colhead{Spectral Type} & \colhead{$\alpha$(2000)} & \colhead{$\delta$(2000)} & \colhead{V (mag)} & \colhead{Nights}
}
\startdata
HD 3384 & SAO 074146 & & G0 &	00:36:55.86	&	+22 18 12.9	& 8.86 [0.02] &	4 \\
BD\textminus07 435 & SAO 129922 & & G0 &	02:27:58.43	&	\textminus07 12 12.7	& 9.74 [0.03] &	4 \\
HD 22319 & SAO 076021 & &	G0 &	03:36:28.81	&	+23 47 50.3	& 8.62 [0.02] &	4 \\
HD 28099 & SAO 093936 & Hyades 64 &	G2V &	04:26:40.12	&	+16 44 48.8	& 8.12 [0.02]	& 25 \\
HD 29461 & SAO 094049 & Hyades 106 &	G5 &	04:38:57.31	&	+14 06 20.1	& 7.95 [0.002] &	32 \\
HD 43965 & SAO 078236 & &	G0 &	06:20:05.02	&	+24 34 00.3	& 7.64 [0.01]	& 6 \\
HD 83789 & SAO 098710 & &	G0 &	09:41:11.49	&	+11 33 25.5	& 8.79 [0.02] &	33 \\
BD+24 2810 & SAO 083619 & &	G0 &	15:01:18.07	&	+23 51 02.8	& 9.33 [0.02] &	12 \\
HD 137272 & SAO 159249 & &	G2/3V &	15:25:32.69	&	\textminus13 44 04.6	&	9.36 [0.02] &	3 \\
HD 137723 & SAO 121010 & &	G0 &	15:27:18.07	&	+09 42 00.3	& 7.93 [0.02] &	3 \\
HD 138278 & SAO 064731 & &	G0 &	15:29:57.63	&	+32 37 07.5	& 8.36 [0.01] &	4 \\
HD 145478 & SAO 121411 & &	G5V &	16:11:06.41	&	+02 54 51.7	& 8.66 [0.01]	& 3 \\
HD 163492 & SAO 141976 & & 	G3V &	17:56:43.12	&	\textminus09 00 53.3	& 8.60 [0.01]	& 3 \\
HD 169359 & SAO 103670 & &	G0 &	18:23:47.06	&	+14 54 27.8	& 7.80 [0.01] &	9 \\
HD 203311 & SAO 164338 & &	G2V &		21:21:51.08	&	\textminus16 16 25.9	& 7.45 [0.01] &	3 \\
HD 217014 & SAO 090896	& 51 Peg &	G2IV	&	22:57:27.98	&	+20 46 07.8	&	5.46 [0.05]	&	14\\
SA 115-271 & \textminus & &	F8 &	23:42:41.82	&	+00 45 13.1	& 9.70 [0.0005] &	16 \\
\enddata
\end{deluxetable*}
\indent Spectra of SAO 66118 showed a distinct change in spectral slope near 1.0 microns in August 2015 that did not appear in July 2015 or earlier in July 2012 (see Figure \ref{fig:rank3}). We exclude SAO 66118 from our list of reliable solar-analog stars. We conclude that the remaining stars in the set are reliable calibration stars and consistent with the solar spectrum at the $<10\%$ level.\\
\indent Following the above analysis of the initial 8 primary solar analogs, we conclude that 2 of the 8 stars in our initial set do not satisfy our criteria for rank 1 solar analogs. Therefore, we have reassessed our original group of primary solar analogs. The stars selected for the list were those that we have both classified as rank 1 and have observed on at least 3 nights. Most stars selected for the expanded list have also been observed over an extensive timespan, typically between 2 and 5 years, in order to verify the star spectrum is repeatable. Our final set of 17 primary solar analogs is presented in Table \ref{tab:newset}. \\
\indent When comparing spectra of the same star taken over several years, we found some inconsistencies that appear to be instrumental sensitivity effects. We used 5 stars with more than 10 separate nights of observation each, and compared them by taking the ratio of each to the overall average spectrum. Figure \ref{fig:anomalies} shows the resulting spectra of 4 of these, with an overall pattern of a dip in relative flux between 1 and 1.8 microns. Over shorter time intervals (weeks to a few months) we find that these stars do not show any change in spectral shape. On individual nights the observations of these stars and other solar-analog stars near the asteroid targets are flat to better than the $\pm10\%$ criterion.\\
\indent Although we do not fully understand the nature of these spectral changes, it is possible that changes in the telescope instrumental sensitivity, or temporary effects due to weather could be responsible. We are confident that the stars are consistent and that intrinsic variability is not likely to be the cause. We know that there were changes in the telescope and instrument configuration, cleaning of the mirror, changing of the dichroic,  and possible ice on the window. However, some specific checks of conditions on particular dates make these seem unlikely. These will not affect the data on a particular night, as they are eliminated by the comparison of the target to the calibration star which both include the same instrument function on a given night.\\ 
\indent Our observing procedure does not use the dichroic and we guide directly with the light from the slit jaws. However, some tests indicate that perhaps the centering of the star and short exposures with few coadded images can lead to changes in spectral slope between 0.8-1.0 microns, but usually less than our 10\% repeatability criterion. We have also checked the influences of airmass on spectral slope and find that there is no particular trend with observing conditions. We have the most spectral slope variation at low airmass, not high airmass. The results of this airmass-spectral slope test are shown in Figure \ref{fig:airmass}.

\subsection{Final Ranking of Candidate Stars}

\indent We compared the remaining 167 candidate stars by dividing each candidate's prism and LXD spectrum by that of at least one of our 17 primary solar analogs, which is typically observed on the same night of the star. The flatness of the resulting ratio spectrum is indicative of how similar the candidate is to the primary solar-analog star, allowing us to meaningfully compare the two stars. The spectral slopes, features, and relative photometry of the resulting comparative spectrum are then assessed using our classification scheme to assign the spectrum its ranking. Each candidate star is assigned its final ranking on a case-by-case basis by considering the rankings of the candidate's spectral comparison to one or more of our 17 primary solar analogs. \\
\indent We present our complete set of 184 solar-type stars, consisting of our final set of 17 primary solar analogs and the remaining 167 candidates, and their assigned rankings in Table \ref{tab:thetable}. Each star's ranking is displayed as both a number (corresponding to the rank) and a colored square in the second-to-last column, with the color of the square indicative of its ranking. As delineated in Section \ref{subsec:scheme}, candidates that are deemed reliable solar analogs without corrections are categorized as rank 1, indicated by a green square in the table. Stars whose prism spectra need polynomial-fit corrections are given a yellow square, and those not correctable due to variability, narrowband nonsolar features, or spectral shape are given a red square. Only the star's ranking in the prism wavelength range is presented because we conclude nearly all of our candidates to be rank 1 in the LXD wavelength range. Each star's spectral type, equatorial coordinates (equinox and epoch J2000.0), V (mag), and V\textminus J, V\textminus H, and V\textminus K colors (in magnitudes) with associated errors are also presented in the table. Any number in the rightmost column corresponds to the footnote describing additional information on a specific star.

\section{Discussion} \label{sec:discussion}
Of our complete set of 184 stars, 145 stars have been classified as reliable solar analogs (rank 1), 21 as needing corrections with polynomial fitting (rank 2), and 18 as not suitable for use as a near-infrared solar standard. The entire set is distributed fairly uniformly in R.A. across the sky, ranging from \textminus27 to +67 degrees in decl. The northern limit is the pointing limit of the IRTF. The overall sky distribution results from the combined effects of our target distribution, weather, and scheduling. Near-Earth asteroids can appear anywhere in the sky when close to the Earth. As our program continues, we will fill in more of the sky as needed to find solar-like stars close to our asteroids. \\
\indent This analysis of the comparison stars also affects the reliability and repeatability of our asteroid spectra. Our choice of the acceptable slope variation for a rank 1 stars of 10\% is somewhat arbitrary, but is based on our experience estimating our own internal consistency and repeatability. Our choice of $\pm 10\%$ is consistent with the $1\sigma$ 4.6\% repeatability for SpeX prism observations reported by \cite{2020ApJS..247...73M}, if we assume that 10\% is about a $2\sigma$ limit, or 95\% confidence level. 
Although we do expect and sometimes see variability in the asteroid spectra, it is generally at a level below 20\% in slope, and usually affects only part of the spectrum (most often the 1.5-2.5 micron region). Our observing program is primarily focused on the thermal emission contribution to the spectrum, which for these near-Earth objects is at least 50\% of the total flux at wavelengths of 3.5 microns and longer. As we noted earlier, the comparison stars are internally consistent on short time-scales of weeks or months. We only see unexplained spectral changes over years, and then mostly at wavelengths shorter than 1.5 microns. Asteroid spectra, particularly S-complex objects, have spectral features in this region that vary by 20\% or more from a flat line, due to pyroxene absorptions or other mineral components (e.g. \cite{2018Icar..303..220H}). The additional uncertainty due to the comparison star spectra does not dominate the result. The asteroid targets are always compared to at least two solar-analog stars, often more, and the final spectra are combined using a weighted average. Our targets are usually observed on at least three separate nights over 2-3 weeks, so the consistency can also be checked on different nights using different nearby stars. Any star that looks like a rank 3 is not included in the analysis. We do not see widespread variability in these measurements but are adjusting our observing techniques to improve the repeatability by doing more coadds for the stars.  

\subsection{Comparing Our Classifications to Other Literature} \label{subsec:rank3cases}

\begin{figure*}
\centerline{\includegraphics[scale = 0.8]{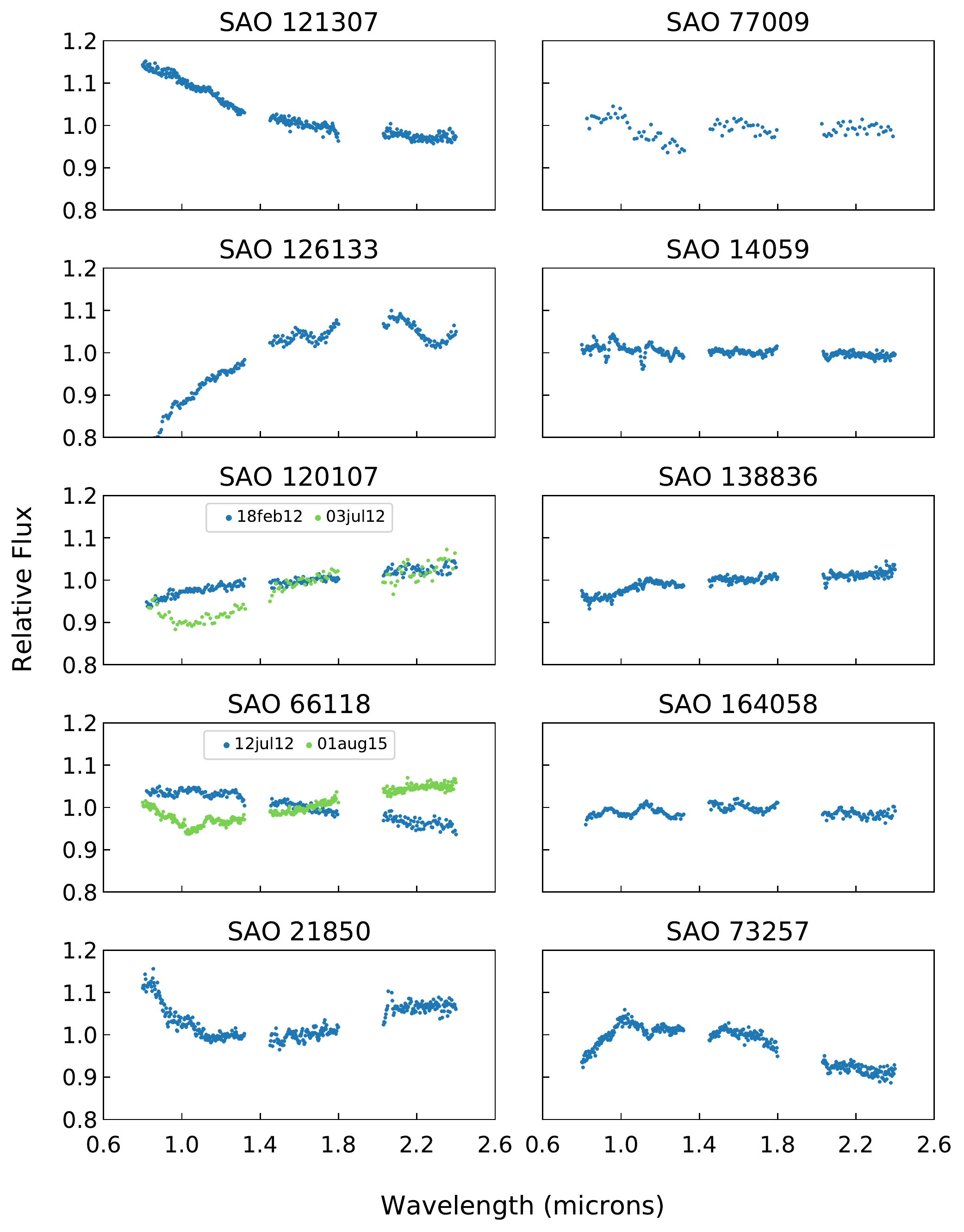}}
\caption{Prism spectra comparing the rank 3 stars discussed in Section \ref{subsec:rank3cases} with one of our 17 solar-analog stars. Stars with only a single spectrum characteristic of the star plotted were assigned rank 3 for having nonsolar features, while those with two spectra plotted show variability on the time scale shown. The telluric water features centered at 1.35 and 1.9 microns have been removed for clarity.}
\label{fig:rank3}
\end{figure*}

\indent \cite{1980A&A....91..221H} identified 4 stars---16 Cyg B and Hyades 64, 106, and 142---as the closest solar spectral analogs, often referred to as Hardorp class 1 in the literature, based on their spectral energy distribution in the 3000-8200 \AA \ range in the visible and the ultraviolet line spectra presented by \cite{1978A&A....63..383H}. Of Hardorp's list, our set includes 16 Cyg B (SAO 31899), Hyades 64 (SAO 93936) and Hyades 106 (SAO 94049), for which we have ample near-infrared observations from 0.7 to 5.3 microns. Similar to the classification of \cite{1980A&A....91..221H}, we classify 16 Cyg B, Hyades 64, and Hyades 106 as rank 1, corresponding to reliable solar analogs by our criteria (note we did not observe Hyades 142). We have also included both Hyades 64 and Hyades 106 in our final list of primary solar analogs (see Section \ref{subsec:analogs}). We observed 16 Cyg B which has been used as a primary standard in near-infrared asteroid studies previously. However, it was often necessary to de-focus the telescope in order to avoid saturation in prism mode. Post-upgrade, although the dynamic range was increased, the minimum exposure time was also increased so that it is impossible to avoid saturation except when the seeing is very bad. We discontinued using this star, although we note that even the de-focused observations have a spectral slope consistent with the overall repeatability of our other standards.\\ 
\indent \cite{1980A&A....91..221H} found the energy distribution of 51 Peg (SAO 90896) to closely resemble that of Hyades 64, which we have classified as a primary solar analog in the near-infrared. 51 Peg has been vetted as a solar analog at visible wavelengths,  which we are in agreement with in the near-infrared.  We observe the star's spectral shape to closely resemble that of Hyades 106, as shown in Figure \ref{fig:starcomp}. We were concerned with potential variability from 51 Peg's 2MASS photometry falling outside of two standard deviations from that of Hyades 106, but quality flags indicate that 51 Peg saturated the field. Using pre-2MASS infrared photometry from Vizier catalog II/225 resulted in the colors of 51 Peg being in agreement with those of Hyades 106. In addition, 51 Peg has been consistently classified between G2V and G5V throughout the past 70 years, as shown by Vizier catalog B/mk \citep{2014yCat....1.2023S}. With 14 observations of 51 Peg, we consequently have classified 51 Peg as both a rank 1 star and one of our primary solar analogs. \\ 
\indent After \cite{1978A&A....63..383H} found its ultraviolet spectrum (3640-4100 \AA) to closely resemble that of the Sun, SAO 66118 has been often used as a solar-analog star for observations in both the visible and in the near-infrared. \cite{2004A&A...418.1089S} included SAO 66118 in their list of top 10 solar analogs based on the resemblance of both its spectrum in the wavelength range 3850-6800 \AA \ and the star's B\textminus V, U\textminus B, and b\textminus y colors to those of the Sun. Although we found the near-infrared spectrum of SAO 66118 to roughly match that of solar analog Hyades 106 in our 2012 observation, our 2015 observation shows that the star's spectral shape has changed considerably over the three-year span. The significant change in spectral shape between the 2012 and 2015 observations is shown in Figure \ref{fig:rank3}. We classify the star as rank 3 by our classification scheme. \cite{2007AJ....133..862H} reported SAO 66118 to be variable, categorizing the star as a high-activity variable, the class corresponding to the most variable stars of their three classes. \\ 
\indent \cite{2005AJ....130.2318K} reported SAO 121307 as potentially comparable as a solar analog to HR 6060, once regarded as the closest ever solar twin. This conclusion was based on high-resolution Keck HIRES spectroscopic data (4475-6900 \AA) in addition to the similarity of the star's effective temperature, luminosity, mass, age, light-metal abundances, and rotational velocity to that of the Sun. Our spectral observations of SAO 121307, however, reveal the spectral shape of the star to increase significantly toward the blue end of the prism spectrum (near 0.8 microns). As shown in Figure \ref{fig:rank3}, where the spectrum of SAO 121307 is compared to that of SAO 98710, one of our primary solar analogs, collected on the same night. The blue end of the comparative prism spectrum exceeds the 10\% maximum deviation limit from a straight line, but remains within a deviation of 20\%. Thus, we classify SAO 121307 as rank 2, or needing polynomial-fit corrections to be used as a solar analog in the prism range of the near-infrared. \\
\indent Since being used as a Pluto comparison star by \cite{1980BAAS...12..729T}, SAO 120107 has been used occasionally as a solar-analog star for asteroid spectral observations, primarily in the near-infrared using the IRTF SpeX instrument. The spectrum of the star closely resembles that of many stars in our list of primary solar analogs for all seven of our observations spanning from June 2009 to February 2012. When we observed SAO 120107 four months later in June 2012, however, the star's spectral shape had changed significantly and needed to be corrected with polynomial fitting, as shown in Figure \ref{fig:rank3}. Although we have ranked SAO 120107 as a rank 3 for variability, the star's spectral shape in our observations is consistently correctable with fitting and thus may be usable for some purposes. We stopped using SAO 120107 as a calibration star after our only anomalous observation in 2012, so further investigation into the star's near-infrared variability may be warranted.  \\
\indent SAO 126133 was included in the set of UBV solar-analog stars presented by \cite{1973AJ.....78..959L} and has since been frequently utilized as a solar analog in asteroid and comet observations in both the visible and in the near-infrared. When we observed the star in September 2008 (shown in Figure \ref{fig:rank3}), we detected  features, many of which are broad and deep, and found the star's spectral flux ratio to decrease rapidly toward the 0.8-micron end. These broad features are centered near 1.6 microns and 2.3 microns, with the latter being the most prominent. As a result of both the spectral shape and features, we report SAO 126133 as a rank 3 star in our list. \\
\indent We have concluded that many of the rank 3 stars in our list are not adequate solar analogs for our purposes from 0.7 to 2.5 microns as a result of possessing low-to-medium resolution, nonsolar spectral features in this wavelength range. The set of stars for which we detect such features consists of SAO 126133, SAO 21850, SAO 77009, SAO 14059, SAO 138836, SAO 164058, and SAO 73257. Representative spectra for each of these stars are shown in Figure \ref{fig:rank3}. Although such features render the stars unsuitable for our asteroid observations, we believe that these features may be of interest for those studying stellar systems, disks, or planetary systems. Although these stars are not suitable for our calibration purposes, these spectral features or variability may be interesting in themselves for other types of study, so we bring them to the attention of other observers.\\

\section{Conclusion} \label{sec:conclusion}
\indent We have presented a list of 184 stars that we have observed as potential solar analogs as indicated by their catalog spectral type and solar-like V\textminus J and V\textminus K colors, shown in Table \ref{tab:thetable}. Based on our classification system, each candidate has been assigned one of three ranks indicative of the star's quality as a solar analog from 0.8 to 2.5 microns, shown as a colored box in the final column of Table \ref{tab:thetable}. Of our set of 184 candidate stars, we conclude 145 to be adequate solar analogs (rank 1), 21 as needing spectral corrections with low-order polynomial fitting (rank 2), and 18 as effectively nonsolar owing to spectral shape, variability, or features at low to medium resolution (rank 3). We conclude that all but 5 of our candidates are sufficient, rank 1 solar analogs in the longer wavelength range from 2.5 to 4.2 microns. However, we do not have observations in this wavelength range for 12 stars. The average colors of the stars classified as rank 1 or rank 2 are V\textminus J=1.148, V\textminus H=1.418, and V\textminus K=1.491, with the entire set being distributed nearly uniformly across the sky from \textminus27 to +67 degrees in decl. We present a set of 17 reliable solar analogs we have classified as rank 1 and have observed for more than 3 nights, prioritizing stars with observations spanning at least 2 years. We have discovered that several stars studied and vetted as solar analogs at visible wavelengths fail to meet our rank 1 or even rank 2 criteria for solar analogs, most frequently owing to variability or possessing nonsolar features; these stars include SAO 66118, SAO 121307, SAO 120107, and SAO 126133. In addition, we have presented the spectral data of stars in which we observe nonsolar features with the hopes of potentially contributing to research outside the purposes of this paper which may be worth further investigation.
 
\acknowledgments

This work includes data collected at the Infrared Telescope Facility, which is operated by the University of Hawaii under contract NNH14CK55B with the National Aeronautics and Space Administration. The authors acknowledge the sacred nature of Maunakea, and appreciate the opportunity to observe from the mountain. We thank an anonymous reviewer for helpful comments that improved this paper. C.D.L. was partially supported by the NASA Space Grant program at the University of Arizona. C.D.L. and E.S.H. were partially supported by the NASA OSIRIS-REx Asteroid Sample Return Mission contract NNM10AA11C - Marshall Space Flight Center. S.E.M. was partially supported by NASA Earth and Space Science Fellowship NNX15AR14H. R.J.V., Y.R.F. and C.M. were partially supported by AST-0808064. R.J.V., Y.R.F., E.S.H., and C.M. were partially supported by AST-1109855. E.S.H. and S.E.M. were partially supported by NNX12AF24G. E.S.H., S.E.M. and J.L.C. were partially supported by NNX13AQ46G. M.L.H. was partially supported by NNA14AB05A, SSERVI Center for Lunar and Asteroid Surface Science.

\begin{longrotatetable}
\begin{deluxetable*}{lllllllllllllcc}
\tablecaption{Rankings for Solar-analog Candidates Sorted by R.A., with V Photometry from the Hipparcos and Tycho Catalogs (Except for HD 217014, Which Is from Vizier Catalog II/225), and JHK Photometry from 2MASS\label{tab:thetable}}
\tablewidth{1600pt}
\tabletypesize{\scriptsize}
\tablehead{ 
 & & &  & &  &  & & &
 \multicolumn{4}{c}{Errors} & & \\ \cline{10-13}
\colhead{Star} & \colhead{SAO Number} & \colhead{Spectral Type} &
\colhead{$\alpha$(2000)} & \colhead{$\delta$(2000)} & 
\colhead{V} & \colhead{V\textminus J} & 
\colhead{V\textminus H} & \colhead{V\textminus K} & 
\colhead{V} & \colhead{V\textminus J} & \colhead{V\textminus H} & \colhead{V\textminus K} & \colhead{Ranking} & \colhead{Notes}
} 
\startdata
\setcounter{footnote}{0}HD 377                   &	SAO 109027	&	G2V	&	00:08:25.75	&	+06 37 00.5	&	7.59	&	1.168	&	1.441	&	1.474	&	0.010 & 0.023 & 0.023 & 0.023	&	1 \hspace{0.2cm}\fcolorbox{black}{green}{\rule{0pt}{3pt}\rule{3pt}{0pt}} & \\
HD 1204                  &	SAO 147192	&	G3V	&	00:16:22.24	&	\textminus19 23 58.6	&	8.77	&	1.100	&	1.403	&	1.455	&	0.010 & 0.021 & 0.022 & 0.023	&	1 \hspace{0.2cm}\fcolorbox{black}{green}{\rule{0pt}{3pt}\rule{3pt}{0pt}} & \\
CD\textminus25 123                &	SAO 166233	&	G0	&	00:23:38.40	&	\textminus24 38 54.6	&	10.01	&	1.177	&	1.498	&	1.551	&	0.020 & 0.033 & 0.050 & 0.030	&	2 \hspace{0.2cm}\fcolorbox{black}{yellow}{\rule{0pt}{3pt}\rule{3pt}{0pt}} & \footnote{\label{LXDnoise}LXD Signal-to-noise too low to adequately characterize}  \\
BD+48 182                &	SAO 036507	&	G0	&	00:36:43.37	&	+48 49 41.8	&	8.67	&	1.179	&	1.486	&	1.556	&	0.010 & 0.026 & 0.020 & 0.022	&	1 \hspace{0.2cm}\fcolorbox{black}{green}{\rule{0pt}{3pt}\rule{3pt}{0pt}} & \\
HD 3384                  &	SAO 074146	&	G0	&	00:36:55.86	&	+22 18 12.9	&	8.86	&	1.119	&	1.364	&	1.452	&	0.020 & 0.028 & 0.034 & 0.026	&	1 \hspace{0.2cm}\fcolorbox{black}{green}{\rule{0pt}{3pt}\rule{3pt}{0pt}} & \\
HD 3964                  &	SAO 128909	&	G5V	&	00:42:09.78	&	\textminus03 03 23.0	&	8.38	&	1.206	&	1.492	&	1.552	&	0.010 & 0.026 & 0.029 & 0.026	&	1 \hspace{0.2cm}\fcolorbox{black}{green}{\rule{0pt}{3pt}\rule{3pt}{0pt}} & \\
HD 5372                  &	SAO 021850	&	G5	&	00:56:17.37	&	+52 29 28.5	&	7.52	&	1.137	&	1.393	&	1.458	&	0.009 & 0.021 & 0.020 & 0.028	&	3 \hspace{0.2cm}\fcolorbox{black}{red}{\rule{0pt}{3pt}\rule{3pt}{0pt}} & \\
HD 7983                  &	SAO 129224	&	G3V	&	01:18:59.99	&	\textminus08 56 22.2	&	8.90	&	1.182	&	1.488	&	1.546	&	0.020 & 0.034 & 0.041 & 0.029	&	2 \hspace{0.2cm}\fcolorbox{black}{yellow}{\rule{0pt}{3pt}\rule{3pt}{0pt}} & \\
HD 7944                  &	SAO 074636	&	G0	&	01:19:26.58	&	+24 24 11.3	&	8.50	&	1.118	&	1.396	&	1.457	&	0.010 & 0.021 & 0.026 & 0.023	&	1 \hspace{0.2cm}\fcolorbox{black}{green}{\rule{0pt}{3pt}\rule{3pt}{0pt}} & \\
HD 8004                  &	SAO 022200	&	G0	&	01:20:36.94	&	+54 57 44.5	&	7.20	&	1.112	&	1.332	&	1.411	&	0.010 & 0.026 & 0.034 & 0.026	&	1 \hspace{0.2cm}\fcolorbox{black}{green}{\rule{0pt}{3pt}\rule{3pt}{0pt}} & \\
HD 8100                  &	SAO 054620	&	G0	&	01:21:03.98	&	+38 02 03.1	&	7.87	&	1.144	&	1.412	&	1.515	&	0.020 & 0.031 & 0.048 & 0.028	&	1 \hspace{0.2cm}\fcolorbox{black}{green}{\rule{0pt}{3pt}\rule{3pt}{0pt}} & \\
HD 9224                  &	SAO 074767	&	G0V	&	01:31:19.52	&	+29 24 47.1	&	7.31	&	1.168	&	1.429	&	1.486	&	0.020 & 0.028 & 0.029 & 0.026	&	3 \hspace{0.2cm}\fcolorbox{black}{red}{\rule{0pt}{3pt}\rule{3pt}{0pt}} & \\
HD 9729                  &	SAO 147888	&	G2V	&	01:35:01.48	&	\textminus12 05 06.6	&	8.62	&	1.115	&	1.394	&	1.470	&	0.010 & 0.026 & 0.029 & 0.028	&	1 \hspace{0.2cm}\fcolorbox{black}{green}{\rule{0pt}{3pt}\rule{3pt}{0pt}} & \\
HD 10785                 &	SAO 147997	&	G1/2V	&	01:45:16.37	&	\textminus15 53 44.4	&	8.50	&	1.140	&	1.396	&	1.484	&	0.020 & 0.029 & 0.039 & 0.030	&	1 \hspace{0.2cm}\fcolorbox{black}{green}{\rule{0pt}{3pt}\rule{3pt}{0pt}} & \\
HD 11532                 &	SAO 110201	&	G8III/IV	&	01:53:18.37	&	+00 22 23.3	&	9.71	&	1.131	&	1.440	&	1.538	&	0.020 & 0.031 & 0.055 & 0.034	&	1 \hspace{0.2cm}\fcolorbox{black}{green}{\rule{0pt}{3pt}\rule{3pt}{0pt}} & \\
HD 11616                 &	SAO 110212	&	G5	&	01:54:11.98	&	+09 57 02.3	&	7.80	&	1.139	&	1.372	&	1.506	&	0.010 & 0.021 & 0.037 & 0.026	&	1 \hspace{0.2cm}\fcolorbox{black}{green}{\rule{0pt}{3pt}\rule{3pt}{0pt}} & \\
HD 12165                 &	SAO 129608	&	G2/3V	&	01:59:27.85	&	\textminus00 15 10.9	&	8.85	&	1.146	&	1.416	&	1.483	&	0.020 & 0.030 & 0.043 & 0.030	&	1 \hspace{0.2cm}\fcolorbox{black}{green}{\rule{0pt}{3pt}\rule{3pt}{0pt}} & \\
HD 13545                 &	SAO 092833	&	G0	&	02:12:29.67	&	+16 42 02.1	&	8.16	&	1.152	&	1.382	&	1.417	&	0.010 & 0.021 & 0.026 & 0.028	&	1 \hspace{0.2cm}\fcolorbox{black}{green}{\rule{0pt}{3pt}\rule{3pt}{0pt}} & \\
HD 13931                 &	SAO 037918	&	G0	&	02:16:47.38	&	+43 46 22.8	&	7.60	&	1.148	&	1.366	&	1.461	&	0.010 & 0.022 & 0.066 & 0.031	&	1 \hspace{0.2cm}\fcolorbox{black}{green}{\rule{0pt}{3pt}\rule{3pt}{0pt}} & \\
BD\textminus07 435                &	SAO 129922	&	G0	&	02:27:58.43	&	\textminus07 12 12.7	&	9.74	&	1.160	&	1.427	&	1.506	&	0.030 & 0.038 & 0.045 & 0.040	&	1 \hspace{0.2cm}\fcolorbox{black}{green}{\rule{0pt}{3pt}\rule{3pt}{0pt}} & \\
HD 15942                 &	SAO 093004	&	G0	&	02:34:03.64	&	+12 10 51.1	&	7.49	&	1.159	&	1.399	&	1.475	&	0.010 & 0.023 & 0.021 & 0.021	&	2 \hspace{0.2cm}\fcolorbox{black}{yellow}{\rule{0pt}{3pt}\rule{3pt}{0pt}} & \footnote{LXD Signal-to-noise too low to adequately characterize} \\
HD 17134                 &	SAO 168012	&	G3V	&	02:44:14.62	&	\textminus25 29 43.4	&	6.96	&	1.138	&	1.431	&	1.493	&	0.010 & 0.026 & 0.034 & 0.023	&	1 \hspace{0.2cm}\fcolorbox{black}{green}{\rule{0pt}{3pt}\rule{3pt}{0pt}} & \footnote{Listed as a spectroscopic binary in SIMBAD}\\
HD 20347                 &	SAO 056320	&	G0	&	03:17:44.02	&	+38 38 21.2	&	7.28	&	1.139	&	1.397	&	1.450	&	0.020 & 0.030 & 0.027 & 0.028	&	2 \hspace{0.2cm}\fcolorbox{black}{yellow}{\rule{0pt}{3pt}\rule{3pt}{0pt}} & \\
HD 20939                 &	SAO 111157	&	G2V	&	03:22:42.17	&	+02 28 07.0	&	8.82	&	1.107	&	1.341	&	1.456	&	0.020 & 0.028 & 0.039 & 0.027	&	1 \hspace{0.2cm}\fcolorbox{black}{green}{\rule{0pt}{3pt}\rule{3pt}{0pt}} & \\
HD 21630                 &	SAO 056504	&	G0	&	03:31:04.08	&	+39 38 40.9	&	8.64	&	1.139	&	1.391	&	1.457	&	0.010 & 0.021 & 0.026 & 0.021	&	1 \hspace{0.2cm}\fcolorbox{black}{green}{\rule{0pt}{3pt}\rule{3pt}{0pt}} & \\
HD 22197                 &	SAO 111270	&	G0	&	03:34:42.86	&	+06 50 35.9	&	9.65	&	1.264	&	1.533	&	1.617	&	0.030 & 0.040 & 0.038 & 0.034	&	1 \hspace{0.2cm}\fcolorbox{black}{green}{\rule{0pt}{3pt}\rule{3pt}{0pt}} & \footnote{LXD Signal-to-noise too low to adequately characterize} \\
HD 22319                 &	SAO 076021	&	G0	&	03:36:28.81	&	+23 47 50.3	&	8.62	&	1.153	&	1.403	&	1.448	&	0.020 & 0.028 & 0.031 & 0.035	&	1 \hspace{0.2cm}\fcolorbox{black}{green}{\rule{0pt}{3pt}\rule{3pt}{0pt}} & \\
HD 232816                &	SAO 024134	&	F5	&	03:38:19.40	&	+52 35 56.7	&	9.00	&	1.154	&	1.438	&	1.490	&	0.009 & 0.022 & 0.019 & 0.022	&	1 \hspace{0.2cm}\fcolorbox{black}{green}{\rule{0pt}{3pt}\rule{3pt}{0pt}} & \\
HD 23111                 &	SAO 111366	&	G0	&	03:42:37.85	&	+05 28 42.6	&	9.15	&	1.119	&	1.412	&	1.502	&	0.020 & 0.036 & 0.043 & 0.027	&	1 \hspace{0.2cm}\fcolorbox{black}{green}{\rule{0pt}{3pt}\rule{3pt}{0pt}} & \\
HD 23050                 &	SAO 039061	&	G2V 	&	03:43:47.70	&	+42 36 12.1	&	7.47	&	1.136	&	1.422	&	1.474	&	0.020 & 0.033 & 0.028 & 0.027	&	1 \hspace{0.2cm}\fcolorbox{black}{green}{\rule{0pt}{3pt}\rule{3pt}{0pt}} & \\
HD 285233                &	SAO 093672	&	G0	&	03:55:21.16	&	+19 22 51.2	&	8.84	&	1.132	&	1.394	&	1.449	&	0.016 & 0.028 & 0.029 & 0.033	&	1 \hspace{0.2cm}\fcolorbox{black}{green}{\rule{0pt}{3pt}\rule{3pt}{0pt}} & \\
HD 279209                &	SAO 056879	&	G0	&	04:00:16.12	&	+37 59 08.3	&	9.46	&	1.121	&	1.431	&	1.509	&	0.020 & 0.027 & 0.028 & 0.030	&	1 \hspace{0.2cm}\fcolorbox{black}{green}{\rule{0pt}{3pt}\rule{3pt}{0pt}} & \\
HD 276024                &	SAO 039243	&	G0	&	04:00:29.40	&	+40 12 09.2	&	8.61	&	1.110	&	1.363	&	1.439	&	0.010 & 0.021 & 0.045 & 0.022	&	1 \hspace{0.2cm}\fcolorbox{black}{green}{\rule{0pt}{3pt}\rule{3pt}{0pt}} & \\
HD 26090                 &	SAO 076473	&	G0V+G5V	&	04:08:54.35	&	+29 11 26.2	&	8.24	&	1.115	&	1.384	&	1.472	&	0.014 & 0.028 & 0.049 & 0.030	&	1 \hspace{0.2cm}\fcolorbox{black}{green}{\rule{0pt}{3pt}\rule{3pt}{0pt}} & \\
HD 279527                &	SAO 057048	&	G0	&	04:11:04.86	&	+35 13 33.0	&	9.22	&	1.157	&	1.413	&	1.497	&	0.020 & 0.036 & 0.045 & 0.027	&	1 \hspace{0.2cm}\fcolorbox{black}{green}{\rule{0pt}{3pt}\rule{3pt}{0pt}} & \\
HD 27486                 &	SAO 131121	&	G2V	&	04:20:11.38	&	\textminus04 29 35.2	&	8.99	&	1.104	&	1.351	&	1.425	&	0.010 & 0.031 & 0.052 & 0.028	&	1 \hspace{0.2cm}\fcolorbox{black}{green}{\rule{0pt}{3pt}\rule{3pt}{0pt}} & \\
HD 27748                 &	SAO 024601	&	G5	&	04:25:54.88	&	+57 29 42.4	&	8.57	&	1.173	&	1.474	&	1.511	&	0.010 & 0.021 & 0.021 & 0.028	&	1 \hspace{0.2cm}\fcolorbox{black}{green}{\rule{0pt}{3pt}\rule{3pt}{0pt}} & \\
HD 28099                 &	SAO 093936	&	G2V	&	04:26:40.12	&	+16 44 48.8	&	8.12	&	1.227	&	1.477	&	1.573	&	0.020 & 0.030 & 0.041 & 0.029	&	1 \hspace{0.2cm}\fcolorbox{black}{green}{\rule{0pt}{3pt}\rule{3pt}{0pt}} & \\
HD 28192                 &	SAO 131211	&	G0V	&	04:26:48.82	&	\textminus01 43 28.6	&	8.07	&	1.109	&	1.385	&	1.452	&	0.010 & 0.022 & 0.037 & 0.026	&	1 \hspace{0.2cm}\fcolorbox{black}{green}{\rule{0pt}{3pt}\rule{3pt}{0pt}} & \\
HD 29461                 &	SAO 094049	&	G5	&	04:38:57.31	&	+14 06 20.1	&	7.95	&	1.131	&	1.425	&	1.502	&	0.002 & 0.037 & 0.029 & 0.021	&	1 \hspace{0.2cm}\fcolorbox{black}{green}{\rule{0pt}{3pt}\rule{3pt}{0pt}} & \\
HD 30625                 &	SAO 131516	&	G3V	&	04:49:19.29	&	\textminus00 52 10.6	&	8.64	&	1.198	&	1.511	&	1.578	&	0.020 & 0.029 & 0.055 & 0.028	&	1 \hspace{0.2cm}\fcolorbox{black}{green}{\rule{0pt}{3pt}\rule{3pt}{0pt}} & \\
HD 30572                 &	SAO 076777	&	G0	&	04:49:48.03	&	+23 23 44.7	&	8.51	&	1.143	&	1.369	&	1.453	&	0.020 & 0.033 & 0.034 & 0.029	&	1 \hspace{0.2cm}\fcolorbox{black}{green}{\rule{0pt}{3pt}\rule{3pt}{0pt}} & \\
HD 32658                 &	SAO 094308	&	G0	&	05:05:26.77	&	+13 48 10.7	&	9.28	&	1.124	&	1.339	&	1.388	&	0.020 & 0.035 & 0.053 & 0.030	&	2 \hspace{0.2cm}\fcolorbox{black}{yellow}{\rule{0pt}{3pt}\rule{3pt}{0pt}} & \\
HD 33366                 &	SAO 077009	&	G5	&	05:10:44.54	&	+25 08 29.4	&	8.46	&	1.142	&	1.429	&	1.485	&	0.020 & 0.027 & 0.034 & 0.027	&	3 \hspace{0.2cm}\fcolorbox{black}{red}{\rule{0pt}{3pt}\rule{3pt}{0pt}} & \footnote{Broad features from 1.0 to 1.7 microns}\\
HD 34031                 &	SAO 077054	&	G0	&	05:15:11.60	&	+20 03 21.9	&	7.72	&	1.237	&	1.538	&	1.623	&	0.020 & 0.035 & 0.076 & 0.030	&	2 \hspace{0.2cm}\fcolorbox{black}{yellow}{\rule{0pt}{3pt}\rule{3pt}{0pt}} & \\
BD\textminus18 1066               &	SAO 150380	&	G0	&	05:21:54.85	&	\textminus18 50 20.9	&	10.15	&	1.212	&	1.522	&	1.597	&	0.026 & 0.032 & 0.053 & 0.035	&	1 \hspace{0.2cm}\fcolorbox{black}{green}{\rule{0pt}{3pt}\rule{3pt}{0pt}} & \\
HD 36108                 &	SAO 170461	&	F9V	&	05:28:21.03	&	\textminus22 26 02.1	&	6.78	&	1.096	&	1.370	&	1.467	&	0.010 & 0.028 & 0.031 & 0.026	&	2 \hspace{0.2cm}\fcolorbox{black}{yellow}{\rule{0pt}{3pt}\rule{3pt}{0pt}} & \\
HD 37685                 &	SAO 113028	&	G0	&	05:40:46.35	&	+09 15 55.6	&	7.94	&	1.116	&	1.427	&	1.487	&	0.010 & 0.029 & 0.033 & 0.023	&	1 \hspace{0.2cm}\fcolorbox{black}{green}{\rule{0pt}{3pt}\rule{3pt}{0pt}} & \\
HD 246128                &	SAO 077384	&	G0	&	05:40:46.46	&	+26 59 53.4	&	9.03	&	1.101	&	1.346	&	1.455	&	0.020 & 0.028 & 0.033 & 0.047	&	1 \hspace{0.2cm}\fcolorbox{black}{green}{\rule{0pt}{3pt}\rule{3pt}{0pt}} & \\
BD\textminus16 1205               &	SAO 150701	&	G0	&	05:41:35.30	&	\textminus16 26 04.3	&	9.03	&	1.122	&	1.427	&	1.480	&	0.010 & 0.031 & 0.039 & 0.026	&	1 \hspace{0.2cm}\fcolorbox{black}{green}{\rule{0pt}{3pt}\rule{3pt}{0pt}} & \\
HD 246629                &	SAO 094791	&	G0	&	05:42:49.64	&	+19 50 50.9	&	9.49	&	1.177	&	1.411	&	1.475	&	0.030 & 0.040 & 0.038 & 0.037	&	1 \hspace{0.2cm}\fcolorbox{black}{green}{\rule{0pt}{3pt}\rule{3pt}{0pt}} & \\
HD 37693                 &	SAO 025339	&	G0	&	05:43:26.85	&	+52 29 19.6	&	7.14	&	1.151	&	1.433	&	1.495	&	0.010 & 0.021 & 0.023 & 0.022	&	1 \hspace{0.2cm}\fcolorbox{black}{green}{\rule{0pt}{3pt}\rule{3pt}{0pt}} & \\
HD 38466                 &	SAO 170778	&	G1V	&	05:45:09.11	&	\textminus22 15 45.8	&	9.32	&	1.094	&	1.373	&	1.456	&	0.020 & 0.034 & 0.068 & 0.034	&	1 \hspace{0.2cm}\fcolorbox{black}{green}{\rule{0pt}{3pt}\rule{3pt}{0pt}} & \\
HD 248712                &	SAO 058544	&	G0	&	05:53:39.77	&	+33 44 15.1	&	8.81	&	1.090	&	1.332	&	1.435	&	0.020 & 0.035 & 0.039 & 0.030	&	3 \hspace{0.2cm}\fcolorbox{black}{red}{\rule{0pt}{3pt}\rule{3pt}{0pt}} & \footnote{Rank 3 in LXD range, spectral slope from 3 to 4 microns}\\
HD 250285                &	SAO 077845	&	G0	&	06:01:18.25	&	+27 16 52.8	&	9.12	&	1.110	&	1.377	&	1.427	&	0.020 & 0.030 & 0.035 & 0.031	&	1 \hspace{0.2cm}\fcolorbox{black}{green}{\rule{0pt}{3pt}\rule{3pt}{0pt}} & \\
HD 41478                 &	SAO 058755	&	G0	&	06:07:10.00	&	+37 09 51.4	&	8.60	&	1.121	&	1.412	&	1.479	&	0.010 & 0.021 & 0.021 & 0.022	&	2 \hspace{0.2cm}\fcolorbox{black}{yellow}{\rule{0pt}{3pt}\rule{3pt}{0pt}} & \\
BD+66 436                &	SAO 013769	&	F2	&	06:15:45.02	&	+66 08 07.4	&	9.03	&	0.830	&	0.884	&	0.993	&	0.020 & 0.030 & 0.064 & 0.035	&	1 \hspace{0.2cm}\fcolorbox{black}{green}{\rule{0pt}{3pt}\rule{3pt}{0pt}} & \\
HD 43965                 &	SAO 078236	&	G0	&	06:20:05.02	&	+24 34 00.3	&	7.64	&	1.133	&	1.409	&	1.463	&	0.010 & 0.029 & 0.023 & 0.031	&	1 \hspace{0.2cm}\fcolorbox{black}{green}{\rule{0pt}{3pt}\rule{3pt}{0pt}} & \\
HD 45580                 &	SAO 095730	&	G0	&	06:29:03.70	&	+17 44 42.8	&	7.62	&	1.156	&	1.415	&	1.492	&	0.010 & 0.028 & 0.020 & 0.022	&	1 \hspace{0.2cm}\fcolorbox{black}{green}{\rule{0pt}{3pt}\rule{3pt}{0pt}} & \\
HD 49158                 &	SAO 013980	&	G0	&	06:51:10.33	&	+62 13 46.3	&	8.67	&	1.127	&	1.398	&	1.486	&	0.010 & 0.025 & 0.022 & 0.020	&	1 \hspace{0.2cm}\fcolorbox{black}{green}{\rule{0pt}{3pt}\rule{3pt}{0pt}} & \\
HD 50694                 &	SAO 096268	&	G0	&	06:54:48.09	&	+11 25 56.0	&	8.08	&	1.121	&	1.382	&	1.463	&	0.010 & 0.028 & 0.037 & 0.028	&	2 \hspace{0.2cm}\fcolorbox{black}{yellow}{\rule{0pt}{3pt}\rule{3pt}{0pt}} & \footnote{Nonlinear spectral shape, smooth but indicative of potential time-variablility that one may want to avoid} \\
HD 51708                 &	SAO 014059	&	G0	&	07:03:46.17	&	+67 27 25.2	&	7.74	&	1.157	&	1.448	&	1.528	&	0.010 & 0.029 & 0.034 & 0.028	&	3 \hspace{0.2cm}\fcolorbox{black}{red}{\rule{0pt}{3pt}\rule{3pt}{0pt}} & \\
HD 53991                 &	SAO 059815	&	G0	&	07:08:51.08	&	+37 31 30.3	&	8.60	&	1.161	&	1.424	&	1.499	&	0.017 & 0.029 & 0.026 & 0.026	&	1 \hspace{0.2cm}\fcolorbox{black}{green}{\rule{0pt}{3pt}\rule{3pt}{0pt}} & \\
HD 60513                 &	SAO 153080	&	G2V	&	07:34:13.16	&	\textminus16 11 16.0	&	6.72	&	1.119	&	1.389	&	1.499	&	0.010 & 0.023 & 0.052 & 0.022	&	1 \hspace{0.2cm}\fcolorbox{black}{green}{\rule{0pt}{3pt}\rule{3pt}{0pt}} & \\
HD 62928                 &	SAO 097231	&	G0	&	07:46:51.37	&	+14 03 19.6	&	8.46	&	1.163	&	1.446	&	1.500	&	0.010 & 0.026 & 0.021 & 0.020	&	1 \hspace{0.2cm}\fcolorbox{black}{green}{\rule{0pt}{3pt}\rule{3pt}{0pt}} & \\
BD+33 1603               &	SAO 060387	&	G0	&	07:51:13.10	&	+32 58 47.6	&	9.13	&	1.130	&	1.385	&	1.492	&	0.020 & 0.034 & 0.043 & 0.027	&	1 \hspace{0.2cm}\fcolorbox{black}{green}{\rule{0pt}{3pt}\rule{3pt}{0pt}} & \footnote{LXD Signal-to-noise too low to adequately characterize} \\
HD 64942                 &	SAO 135272	&	G3/5V	&	07:55:58.23	&	\textminus09 47 49.9	&	8.34	&	1.095	&	1.368	&	1.469	&	0.010 & 0.021 & 0.034 & 0.025	&	1 \hspace{0.2cm}\fcolorbox{black}{green}{\rule{0pt}{3pt}\rule{3pt}{0pt}} & \\
HD 69270                 &	SAO 135700	&	G3V	&	08:16:06.58	&	\textminus05 14 32.0	&	9.35	&	1.108	&	1.350	&	1.449	&	0.020 & 0.029 & 0.060 & 0.029	&	1 \hspace{0.2cm}\fcolorbox{black}{green}{\rule{0pt}{3pt}\rule{3pt}{0pt}} & \\
HD 71848                 &	SAO 097859	&	G0	&	08:29:55.73	&	+10 28 14.8	&	8.03	&	1.109	&	1.373	&	1.438	&	0.010 & 0.022 & 0.037 & 0.025	&	1 \hspace{0.2cm}\fcolorbox{black}{green}{\rule{0pt}{3pt}\rule{3pt}{0pt}} & \\
HD 72892                 &	SAO 154423	&	G5V	&	08:34:52.59	&	\textminus14 27 24.1	&	8.79	&	1.184	&	1.467	&	1.574	&	0.020 & 0.029 & 0.039 & 0.028	&	1 \hspace{0.2cm}\fcolorbox{black}{green}{\rule{0pt}{3pt}\rule{3pt}{0pt}} & \\
HD 73510                 &	SAO 080324	&	G5	&	08:39:37.12	&	+24 37 13.6	&	8.89	&	1.119	&	1.364	&	1.436	&	0.020 & 0.028 & 0.026 & 0.025	&	1 \hspace{0.2cm}\fcolorbox{black}{green}{\rule{0pt}{3pt}\rule{3pt}{0pt}} & \\
HD 76151                 &	SAO 136389	&	G2V	&	08:54:17.95	&	\textminus05 26 04.0	&	6.00	&	1.129	&	1.470	&	1.544	&	0.020 & 0.042 & 0.028 & 0.030	&	1 \hspace{0.2cm}\fcolorbox{black}{green}{\rule{0pt}{3pt}\rule{3pt}{0pt}} & \\
BD+26 1904               &	SAO 080675	&	G0	&	09:10:25.85	&	+25 48 58.6	&	10.09	&	1.114	&	1.358	&	1.453	&	0.040 & 0.044 & 0.043 & 0.043	&	2 \hspace{0.2cm}\fcolorbox{black}{yellow}{\rule{0pt}{3pt}\rule{3pt}{0pt}} & \\
HD 79282                 &	SAO 061342	&	G5	&	09:14:14.56	&	+33 49 00.9	&	8.28	&	1.114	&	1.359	&	1.453	&	0.010 & 0.026 & 0.048 & 0.028	&	1 \hspace{0.2cm}\fcolorbox{black}{green}{\rule{0pt}{3pt}\rule{3pt}{0pt}} & \\
HD 83789                 &	SAO 098710	&	G0	&	09:41:11.49	&	+11 33 25.5	&	8.79	&	1.176	&	1.450	&	1.549	&	0.020 & 0.038 & 0.045 & 0.035	&	1 \hspace{0.2cm}\fcolorbox{black}{green}{\rule{0pt}{3pt}\rule{3pt}{0pt}} & \\
HD 86811                 &	SAO 061817	&	F8	&	10:01:55.85	&	+37 44 36.1	&	8.95	&	0.952	&	1.156	&	1.226	&	0.020 & 0.029 & 0.026 & 0.024	&	2 \hspace{0.2cm}\fcolorbox{black}{yellow}{\rule{0pt}{3pt}\rule{3pt}{0pt}} & \\
HD 90681                 &	SAO 062060	&	G0	&	10:28:51.39	&	+34 53 08.4	&	7.82	&	1.114	&	1.369	&	1.400	&	0.010 & 0.026 & 0.039 & 0.022	&	2 \hspace{0.2cm}\fcolorbox{black}{yellow}{\rule{0pt}{3pt}\rule{3pt}{0pt}} & \\
HD 91768                 &	SAO 062148	&	G5	&	10:36:22.56	&	+35 07 20.0	&	8.84	&	1.158	&	1.448	&	1.500	&	0.009 & 0.025 & 0.020 & 0.023	&	1 \hspace{0.2cm}\fcolorbox{black}{green}{\rule{0pt}{3pt}\rule{3pt}{0pt}} & \\
HD 95311                 &	SAO 081617	&	G0	&	11:00:36.82	&	+23 42 21.9	&	8.68	&	1.128	&	1.419	&	1.468	&	0.010 & 0.028 & 0.033 & 0.025	&	1 \hspace{0.2cm}\fcolorbox{black}{green}{\rule{0pt}{3pt}\rule{3pt}{0pt}} & \\
HD 100022                &	SAO 156720	&	G2V	&	11:30:26.09	&	\textminus15 19 19.7	&	9.39	&	1.128	&	1.369	&	1.452	&	0.055 & 0.059 & 0.067 & 0.059	&	1 \hspace{0.2cm}\fcolorbox{black}{green}{\rule{0pt}{3pt}\rule{3pt}{0pt}} & \\
BD+07 2471               &	SAO 118998	&	G5	&	11:41:57.34	&	+06 48 24.6	&	9.89	&	1.124	&	1.415	&	1.499	&	0.040 & 0.044 & 0.048 & 0.047	&	1 \hspace{0.2cm}\fcolorbox{black}{green}{\rule{0pt}{3pt}\rule{3pt}{0pt}} & \\
HD 103549                &	SAO 138483	&	G2/3V	&	11:55:23.69	&	\textminus03 35 58.1	&	8.77	&	1.102	&	1.368	&	1.470	&	0.020 & 0.034 & 0.050 & 0.029	&	1 \hspace{0.2cm}\fcolorbox{black}{green}{\rule{0pt}{3pt}\rule{3pt}{0pt}} & \\
HD 104516                &	SAO 180396	&	G0V	&	12:02:13.46	&	\textminus27 38 52.4	&	8.85	&	1.107	&	1.350	&	1.454	&	0.020 & 0.030 & 0.050 & 0.026	&	1 \hspace{0.2cm}\fcolorbox{black}{green}{\rule{0pt}{3pt}\rule{3pt}{0pt}} & \\
HD 104924                &	SAO 119208	&	G0	&	12:04:53.51	&	+09 10 20.5	&	9.41	&	1.058	&	1.325	&	1.397	&	0.020 & 0.039 & 0.057 & 0.030	&	1 \hspace{0.2cm}\fcolorbox{black}{green}{\rule{0pt}{3pt}\rule{3pt}{0pt}} & \\
HD 105901                &	SAO 138636	&	G3/5V	&	12:11:17.86	&	\textminus05 55 33.9	&	8.18	&	1.123	&	1.382	&	1.484	&	0.020 & 0.036 & 0.035 & 0.030	&	1 \hspace{0.2cm}\fcolorbox{black}{green}{\rule{0pt}{3pt}\rule{3pt}{0pt}} & \\
BD+27 2103               &	SAO 082194	&	G0	&	12:13:30.59	&	+27 02 36.5	&	9.90	&	1.199	&	1.469	&	1.533	&	0.020 & 0.028 & 0.031 & 0.027	&	1 \hspace{0.2cm}\fcolorbox{black}{green}{\rule{0pt}{3pt}\rule{3pt}{0pt}} & \\
HD 109098                &	SAO 138836	&	G3/5V 	&	12:32:04.45	&	\textminus01 46 20.5	&	7.31	&	1.140	&	1.384	&	1.507	&	0.010 & 0.031 & 0.034 & 0.018	&	3 \hspace{0.2cm}\fcolorbox{black}{red}{\rule{0pt}{3pt}\rule{3pt}{0pt}} & \\
BD+39 2571               &	SAO 063224	&	G5	&	12:51:40.44	&	+38 22 06.8	&	10.3	&	1.220	&	1.558	&	1.608	&	0.020 & 0.045 & 0.034 & 0.029	&	1 \hspace{0.2cm}\fcolorbox{black}{green}{\rule{0pt}{3pt}\rule{3pt}{0pt}} & \footnote{LXD Signal-to-noise too low to adequately characterize}\\
HD 115269                &	SAO 028691	&	G0	&	13:15:19.20	&	+52 16 40.3	&	9.05	&	1.156	&	1.433	&	1.504	&	0.009 & 0.025 & 0.034 & 0.022	&	3 \hspace{0.2cm}\fcolorbox{black}{red}{\rule{0pt}{3pt}\rule{3pt}{0pt}} & \\
HD 118034                &	SAO 100622	&	G0	&	13:33:57.66	&	+17 28 05.0	&	8.89	&	1.134	&	1.375	&	1.458	&	0.010 & 0.022 & 0.028 & 0.021	&	1 \hspace{0.2cm}\fcolorbox{black}{green}{\rule{0pt}{3pt}\rule{3pt}{0pt}} & \footnote{LXD Signal-to-noise too low to adequately characterize}\\
HD 119856                &	SAO 204825	&	G1V	&	13:46:17.64	&	\textminus30 28 28.1	&	8.21	&	1.130	&	1.376	&	1.466	&	0.010 & 0.026 & 0.054 & 0.028	&	1 \hspace{0.2cm}\fcolorbox{black}{green}{\rule{0pt}{3pt}\rule{3pt}{0pt}} & \\
HD 120050                &	SAO 120107	&	G5III	&	13:46:57.91	&	+06 01 37.7	&	9.26	&	1.204	&	1.481	&	1.542	&	0.020 & 0.033 & 0.043 & 0.048	&	3 \hspace{0.2cm}\fcolorbox{black}{red}{\rule{0pt}{3pt}\rule{3pt}{0pt}} & \footnote{Non-repeatability in spectra---spectral shape changed significantly between February 2012 and June 2012. Further investigation encouraged.}\\
BD+66 844                &	SAO 016368	&	G5	&	14:23:40.91	&	+65 23 43.8	&	9.29	&	1.144	&	1.418	&	1.482	&	0.020 & 0.029 & 0.030 & 0.028	&	1 \hspace{0.2cm}\fcolorbox{black}{green}{\rule{0pt}{3pt}\rule{3pt}{0pt}} & \\
BD+07 2790               &	SAO 120476	&	G0	&	14:27:55.15	&	+06 57 05.9	&	9.24	&	1.087	&	1.321	&	1.469	&	0.020 & 0.035 & 0.059 & 0.030	&	1 \hspace{0.2cm}\fcolorbox{black}{green}{\rule{0pt}{3pt}\rule{3pt}{0pt}} & \\
HD 129171                &	SAO 064262	&	G0	&	14:40:18.39	&	+30 26 37.8	&	7.69	&	1.114	&	1.371	&	1.424	&	0.010 & 0.022 & 0.026 & 0.026	&	1 \hspace{0.2cm}\fcolorbox{black}{green}{\rule{0pt}{3pt}\rule{3pt}{0pt}} & \\
BD+24 2757               &	SAO 083468	&	G0	&	14:41:18.10	&	+23 43 19.6	&	9.01	&	1.138	&	1.397	&	1.465	&	0.019 & 0.031 & 0.038 & 0.030	&	1 \hspace{0.2cm}\fcolorbox{black}{green}{\rule{0pt}{3pt}\rule{3pt}{0pt}} & \\
HD 131526                &	SAO 045252	&	G0	&	14:52:20.91	&	+48 40 14.7	&	7.64	&	1.175	&	1.442	&	1.480	&	0.010 & 0.023 & 0.023 & 0.021	&	1 \hspace{0.2cm}\fcolorbox{black}{green}{\rule{0pt}{3pt}\rule{3pt}{0pt}} & \footnote{LXD Signal-to-noise too low to adequately characterize}\\
HD 131715                &	SAO 140224	&	F8V	&	14:55:16.18	&	\textminus02 06 50.7	&	8.95	&	1.106	&	1.417	&	1.483	&	0.020 & 0.030 & 0.047 & 0.029	&	1 \hspace{0.2cm}\fcolorbox{black}{green}{\rule{0pt}{3pt}\rule{3pt}{0pt}} & \footnote{LXD Signal-to-noise too low to adequately characterize}\\
HD 131790                &	SAO 158903	&	G0V	&	14:56:02.54	&	\textminus15 39 23.5	&	8.00	&	1.155	&	1.347	&	1.476	&	0.010 & 0.023 & 0.028 & 0.021	&	3 \hspace{0.2cm}\fcolorbox{black}{red}{\rule{0pt}{3pt}\rule{3pt}{0pt}} & \\
BD+24 2810               &	SAO 083619	&	G0	&	15:01:18.07	&	+23 51 02.8	&	9.33	&	1.144	&	1.462	&	1.508	&	0.020 & 0.044 & 0.031 & 0.027	&	1 \hspace{0.2cm}\fcolorbox{black}{green}{\rule{0pt}{3pt}\rule{3pt}{0pt}} & \\
HD 134533                &	SAO 101410	&	G0	&	15:10:03.01	&	+12 20 38.7	&	9.34	&	1.151	&	1.408	&	1.472	&	0.020 & 0.029 & 0.034 & 0.029	&	1 \hspace{0.2cm}\fcolorbox{black}{green}{\rule{0pt}{3pt}\rule{3pt}{0pt}} & \\
BD\textminus08 3922               &	SAO 140388	&	G0	&	15:13:44.73	&	\textminus08 37 12.0	&	9.61	&	1.265	&	1.582	&	1.597	&	0.105 & 0.109 & 0.113 & 0.109	&	1 \hspace{0.2cm}\fcolorbox{black}{green}{\rule{0pt}{3pt}\rule{3pt}{0pt}} & \\
HD 137272                &	SAO 159249	&	G2/3V	&	15:25:32.69	&	\textminus13 44 04.6	&	9.36	&	1.170	&	1.447	&	1.508	&	0.020 & 0.039 & 0.045 & 0.031	&	1 \hspace{0.2cm}\fcolorbox{black}{green}{\rule{0pt}{3pt}\rule{3pt}{0pt}} & \\
HD 137723                &	SAO 121010	&	G0	&	15:27:18.07	&	+09 42 00.3	&	7.93	&	1.098	&	1.325	&	1.446	&	0.020 & 0.027 & 0.043 & 0.028	&	1 \hspace{0.2cm}\fcolorbox{black}{green}{\rule{0pt}{3pt}\rule{3pt}{0pt}} & \\
HD 137775                &	SAO 101565	&	G0	&	15:27:35.07	&	+09 53 14.6	&	8.99	&	1.097	&	1.379	&	1.448	&	0.020 & 0.031 & 0.030 & 0.030	&	1 \hspace{0.2cm}\fcolorbox{black}{green}{\rule{0pt}{3pt}\rule{3pt}{0pt}} & \\
HD 138278                &	SAO 064731	&	G0	&	15:29:57.63	&	+32 37 07.5	&	8.36	&	1.168	&	1.392	&	1.458	&	0.010 & 0.021 & 0.019 & 0.019	&	1 \hspace{0.2cm}\fcolorbox{black}{green}{\rule{0pt}{3pt}\rule{3pt}{0pt}} & \\
HD 138573                &	SAO 101603	&	G5IV\textminus V	&	15:32:43.65	&	+10 58 05.9	&	7.21	&	1.183	&	1.468	&	1.548	&	0.010 & 0.023 & 0.028 & 0.020	&	1 \hspace{0.2cm}\fcolorbox{black}{green}{\rule{0pt}{3pt}\rule{3pt}{0pt}} & \\
HD 139287                &	SAO 121093	&	G2/3V	&	15:37:18.15	&	\textminus00 09 49.7	&	8.44	&	1.225	&	1.488	&	1.610	&	0.020 & 0.030 & 0.043 & 0.030	&	1 \hspace{0.2cm}\fcolorbox{black}{green}{\rule{0pt}{3pt}\rule{3pt}{0pt}} & \\
HD 141715                &	SAO 121216	&	G3V	&	15:50:26.06	&	+01 49 08.3	&	8.28	&	1.138	&	1.430	&	1.538	&	0.010 & 0.025 & 0.050 & 0.025	&	1 \hspace{0.2cm}\fcolorbox{black}{green}{\rule{0pt}{3pt}\rule{3pt}{0pt}} & \\
HD 143436                &	SAO 121307	&	G3V	&	16:00:18.84	&	+00 08 13.2	&	8.03	&	1.146	&	1.381	&	1.489	&	0.010 & 0.021 & 0.026 & 0.021	&	3 \hspace{0.2cm}\fcolorbox{black}{red}{\rule{0pt}{3pt}\rule{3pt}{0pt}} & \footnote{Spectral shape increases significantly ($\approx 15\%$) toward the blue end of the prism spectrum (near 0.8 microns)}\\
HD 144873                &	SAO 065083	&	G5	&	16:06:40.00	&	+34 06 10.5	&	8.54	&	1.241	&	1.547	&	1.628	&	0.010 & 0.022 & 0.022 & 0.021	&	2 \hspace{0.2cm}\fcolorbox{black}{yellow}{\rule{0pt}{3pt}\rule{3pt}{0pt}} & \\
HD 145478                &	SAO 121411	&	G5V	&	16:11:06.41	&	+02 54 51.7	&	8.66	&	1.165	&	1.398	&	1.492	&	0.010 & 0.022 & 0.031 & 0.023	&	1 \hspace{0.2cm}\fcolorbox{black}{green}{\rule{0pt}{3pt}\rule{3pt}{0pt}} & \\
HD 146070                &	SAO 184262	&	G1V	&	16:15:19.09	&	\textminus27 12 36.9	&	7.54	&	1.132	&	1.431	&	1.498	&	0.010 & 0.023 & 0.022 & 0.022	&	1 \hspace{0.2cm}\fcolorbox{black}{green}{\rule{0pt}{3pt}\rule{3pt}{0pt}} & \\
HD 153227                &	SAO 121963	&	G3/5V	&	16:58:00.44	&	+02 20 31.1	&	9.81	&	1.071	&	1.333	&	1.390	&	0.030 & 0.036 & 0.043 & 0.036	&	2 \hspace{0.2cm}\fcolorbox{black}{yellow}{\rule{0pt}{3pt}\rule{3pt}{0pt}} & \\
HD 153631                &	SAO 160227	&	G0V	&	17:01:10.76	&	\textminus13 34 01.7	&	7.14	&	1.176	&	1.477	&	1.560	&	0.020 & 0.030 & 0.031 & 0.029	&	1 \hspace{0.2cm}\fcolorbox{black}{green}{\rule{0pt}{3pt}\rule{3pt}{0pt}} & \\
HD 153994                &	SAO 102542	&	G0	&	17:02:21.38	&	+12 24 50.3	&	9.51	&	1.080	&	1.323	&	1.379	&	0.020 & 0.030 & 0.030 & 0.027	&	1 \hspace{0.2cm}\fcolorbox{black}{green}{\rule{0pt}{3pt}\rule{3pt}{0pt}} & \\
HD 154064                &	SAO 102550	&	G5	&	17:02:42.73	&	+12 56 32.2	&	8.33	&	1.177	&	1.463	&	1.499	&	0.010 & 0.029 & 0.023 & 0.019	&	2 \hspace{0.2cm}\fcolorbox{black}{yellow}{\rule{0pt}{3pt}\rule{3pt}{0pt}} & \\
HD 234382                &	SAO 030231	&	G0	&	17:04:30.40	&	+52 09 42.1	&	8.59	&	1.130	&	1.395	&	1.472	&	0.010 & 0.026 & 0.020 & 0.020	&	1 \hspace{0.2cm}\fcolorbox{black}{green}{\rule{0pt}{3pt}\rule{3pt}{0pt}} & \\
HD 159222                &	SAO 066118	&	G1V	&	17:32:00.99	&	+34 16 16.1	&	6.56	&	1.218	&	1.484	&	1.562	&	0.020 & 0.029 & 0.034 & 0.026	&	3 \hspace{0.2cm}\fcolorbox{black}{red}{\rule{0pt}{3pt}\rule{3pt}{0pt}} & \footnote{Spectral changes from 2012 to 2015 indicate non-repeatibility}\\
HD 159333                &	SAO 122512	&	G0	&	17:33:52.82	&	+08 06 13.6	&	8.88	&	2.111	&	2.428	&	2.496	&	0.054 & 0.060 & 0.058 & 0.056	&	1 \hspace{0.2cm}\fcolorbox{black}{green}{\rule{0pt}{3pt}\rule{3pt}{0pt}} & \\
HD 162209                &	SAO 066346	&	G0	&	17:48:13.08	&	+38 13 57.3	&	7.77	&	1.107	&	1.401	&	1.455	&	0.010 & 0.035 & 0.019 & 0.021	&	1 \hspace{0.2cm}\fcolorbox{black}{green}{\rule{0pt}{3pt}\rule{3pt}{0pt}} & \\
HD 163492                &	SAO 141976	&	G3V 	&	17:56:43.12	&	\textminus09 00 53.3	&	8.60	&	1.126	&	1.401	&	1.494	&	0.010 & 0.034 & 0.058 & 0.031	&	1 \hspace{0.2cm}\fcolorbox{black}{green}{\rule{0pt}{3pt}\rule{3pt}{0pt}} & \\
HD 165290                &	SAO 186276	&	G1V	&	18:06:17.29	&	\textminus26 17 02.7	&	9.04	&	1.185	&	1.369	&	1.492	&	0.020 & 0.029 & 0.035 & 0.028	&	2 \hspace{0.2cm}\fcolorbox{black}{yellow}{\rule{0pt}{3pt}\rule{3pt}{0pt}} & \\
HD 165672                &	SAO 123130	&	G5	&	18:06:48.81	&	+06 24 38.0	&	7.77	&	1.143	&	1.419	&	1.465	&	0.010 & 0.022 & 0.034 & 0.026	&	1 \hspace{0.2cm}\fcolorbox{black}{green}{\rule{0pt}{3pt}\rule{3pt}{0pt}} & \\
HD 348088                &	SAO 085831	&	G0	&	18:13:06.93	&	+20 19 34.3	&	8.91	&	1.101	&	1.398	&	1.439	&	0.010 & 0.026 & 0.035 & 0.019	&	1 \hspace{0.2cm}\fcolorbox{black}{green}{\rule{0pt}{3pt}\rule{3pt}{0pt}} & \\
HD 167065                &	SAO 123264	&	G0	&	18:13:18.79	&	+09 05 49.3	&	8.02	&	1.164	&	1.410	&	1.497	&	0.010 & 0.028 & 0.028 & 0.029	&	1 \hspace{0.2cm}\fcolorbox{black}{green}{\rule{0pt}{3pt}\rule{3pt}{0pt}} & \\
HD 169359                &	SAO 103670	&	G0	&	18:23:47.06	&	+14 54 27.8	&	7.80	&	1.148	&	1.397	&	1.461	&	0.010 & 0.021 & 0.031 & 0.020	&	1 \hspace{0.2cm}\fcolorbox{black}{green}{\rule{0pt}{3pt}\rule{3pt}{0pt}} & \\
BD+28 2993               &	SAO 086008	&	G0	&	18:24:11.62	&	+28 17 25.1	&	9.17	&	1.173	&	1.439	&	1.479	&	0.010 & 0.021 & 0.020 & 0.023	&	1 \hspace{0.2cm}\fcolorbox{black}{green}{\rule{0pt}{3pt}\rule{3pt}{0pt}} & \\
HD 170331                &	SAO 186884	&	G5V	&	18:29:52.02	&	\textminus26 01 31.4	&	8.81	&	1.168	&	1.403	&	1.523	&	0.020 & 0.030 & 0.029 & 0.029	&	1 \hspace{0.2cm}\fcolorbox{black}{green}{\rule{0pt}{3pt}\rule{3pt}{0pt}} & \\
BD+35 3269               &	SAO 067043	&	G5	&	18:30:25.63	&	+35 43 39.1	&	9.13	&	1.175	&	1.463	&	1.539	&	0.020 & 0.028 & 0.026 & 0.030	&	2 \hspace{0.2cm}\fcolorbox{black}{yellow}{\rule{0pt}{3pt}\rule{3pt}{0pt}} & \\
SA 110\textminus361            & \textminus	&	N/A	&	18:42:45.01	& +00 08 04.7	&	12.43	&	1.214	& 1.499	&	1.565	&	0.002 & 0.010 & 0.022 & 0.043	&	1 \hspace{0.2cm}\fcolorbox{black}{green}{\rule{0pt}{3pt}\rule{3pt}{0pt}} & \footnote{Spectral type not listed on SIMBAD}\\
HD 174466                &	SAO 161908	&	G2V	&	18:51:15.80	&	\textminus16 09 42.5	&	8.81	&	1.661	&	1.923	&	1.996	&	0.034 & 0.039 & 0.047 & 0.039	&	1 \hspace{0.2cm}\fcolorbox{black}{green}{\rule{0pt}{3pt}\rule{3pt}{0pt}} & \\
HD 175179                &	SAO 142780	&	G5V	&	18:54:23.20	&	\textminus04 36 18.6	&	9.03	&	1.101	&	1.400	&	1.487	&	0.020 & 0.030 & 0.050 & 0.034	&	3 \hspace{0.2cm}\fcolorbox{black}{red}{\rule{0pt}{3pt}\rule{3pt}{0pt}} & \\
HD 175702                &	SAO 104268	&	G0	&	18:56:05.80	&	+15 21 56.4	&	7.67	&	1.175	&	1.407	&	1.463	&	0.010 & 0.021 & 0.022 & 0.019	&	1 \hspace{0.2cm}\fcolorbox{black}{green}{\rule{0pt}{3pt}\rule{3pt}{0pt}} & \\
HD 176972                &	SAO 104383	&	G5	&	19:01:53.16	&	+19 10 11.7	&	7.88	&	1.139	&	1.434	&	1.465	&	0.010 & 0.028 & 0.025 & 0.019	&	2 \hspace{0.2cm}\fcolorbox{black}{yellow}{\rule{0pt}{3pt}\rule{3pt}{0pt}} & \\
HD 177780                &	SAO 048053	&	G3V	&	19:04:16.38	&	+41 00 11.4	&	8.37	&	1.149	&	1.395	&	1.439	&	0.010 & 0.021 & 0.029 & 0.025	&	1 \hspace{0.2cm}\fcolorbox{black}{green}{\rule{0pt}{3pt}\rule{3pt}{0pt}} & \\
HD 231043                &	SAO 104675	&	G0	&	19:16:38.92	&	+16 40 04.9	&	9.25	&	1.173	&	1.389	&	1.460	&	0.020 & 0.029 & 0.041 & 0.033	&	2 \hspace{0.2cm}\fcolorbox{black}{yellow}{\rule{0pt}{3pt}\rule{3pt}{0pt}} & \\
HD 183542                &	SAO 162700	&	G2/3V	&	19:30:35.74	&	\textminus11 33 43.9	&	9.71	&	1.182	&	1.478	&	1.570	&	0.030 & 0.036 & 0.056 & 0.040	&	1 \hspace{0.2cm}\fcolorbox{black}{green}{\rule{0pt}{3pt}\rule{3pt}{0pt}} & \\
HD 184403                &	SAO 087353	&	G0	&	19:33:26.21	&	+23 29 51.0	&	7.81	&	1.249	&	1.489	&	1.630	&	0.010 & 0.028 & 0.083 & 0.021	&	2 \hspace{0.2cm}\fcolorbox{black}{yellow}{\rule{0pt}{3pt}\rule{3pt}{0pt}} & \\
HD 186427               &	SAO 031899	&	G3V	&	19:41:51.97	&	+50 31 03.1	&	6.20	&	1.210	&	1.600	&	1.549	&	0.020 & 0.028 & 0.028 & 0.026	&	1 \hspace{0.2cm}\fcolorbox{black}{green}{\rule{0pt}{3pt}\rule{3pt}{0pt}} & \\
HD 186413                &	SAO 124998	&	G3V	&	19:44:04.39	&	+03 30 27.8	&	7.99	&	1.112	&	1.377	&	1.446	&	0.010 & 0.025 & 0.028 & 0.026	&	1 \hspace{0.2cm}\fcolorbox{black}{green}{\rule{0pt}{3pt}\rule{3pt}{0pt}} & \\
HD 186932                &	SAO 105240	&	G0	&	19:46:37.82	&	+17 48 10.5	&	8.10	&	1.173	&	1.426	&	1.458	&	0.010 & 0.025 & 0.022 & 9.995	&	1 \hspace{0.2cm}\fcolorbox{black}{green}{\rule{0pt}{3pt}\rule{3pt}{0pt}} & \\
HD 187876                &	SAO 032031	&	G0	&	19:49:12.18	&	+57 24 33.6	&	7.76	&	1.063	&	1.340	&	1.426	&	0.020 & 0.036 & 0.060 & 0.028	&	1 \hspace{0.2cm}\fcolorbox{black}{green}{\rule{0pt}{3pt}\rule{3pt}{0pt}} & \\
HD 187897                &	SAO 125154	&	G5	&	19:52:09.39	&	+07 27 36.2	&	7.13	&	1.074	&	1.384	&	1.449	&	0.010 & 0.022 & 0.023 & 0.023	&	1 \hspace{0.2cm}\fcolorbox{black}{green}{\rule{0pt}{3pt}\rule{3pt}{0pt}} & \\
HD 190524                &	SAO 163258	&	G3V	&	20:05:48.70	&	\textminus15 45 22.4	&	8.44	&	1.137	&	1.403	&	1.472	&	0.020 & 0.028 & 0.035 & 0.027	&	1 \hspace{0.2cm}\fcolorbox{black}{green}{\rule{0pt}{3pt}\rule{3pt}{0pt}} & \\
HD 346383                &	SAO 088564	&	G0	&	20:21:25.03	&	+23 31 15.6	&	8.87	&	1.153	&	1.391	&	1.452	&	0.020 & 0.033 & 0.039 & 0.029	&	1 \hspace{0.2cm}\fcolorbox{black}{green}{\rule{0pt}{3pt}\rule{3pt}{0pt}} & \\
HD 196361                &	SAO 070242	&	G5	&	20:35:38.59	&	+36 28 29.7	&	8.24	&	1.166	&	1.431	&	1.483	&	0.010 & 0.023 & 0.022 & 0.019	&	1 \hspace{0.2cm}\fcolorbox{black}{green}{\rule{0pt}{3pt}\rule{3pt}{0pt}} & \\
HD 197195                &	SAO 106391	&	G5	&	20:41:53.31	&	+12 58 49.6	&	8.24	&	1.060	&	1.287	&	1.341	&	0.010 & 0.025 & 0.035 & 0.019	&	1 \hspace{0.2cm}\fcolorbox{black}{green}{\rule{0pt}{3pt}\rule{3pt}{0pt}} & \\
BD\textminus00 4074               &	SAO 126133	&	F8	&	20:43:11.96	&	+00 26 13.1	&	9.90	&	1.073	&	1.321	&	1.436	&	0.020 & 0.027 & 0.064 & 0.030	&	3 \hspace{0.2cm}\fcolorbox{black}{red}{\rule{0pt}{3pt}\rule{3pt}{0pt}} & \footnote{Ample features, many of which are broad and deep. Prism spectrum decreases rapidly toward the blue end (near 0.8 microns).}\\
HD 198176                &	SAO 189737	&	G2V	&	20:49:24.55	&	\textminus26 56 14.7	&	8.73	&	1.117	&	1.450	&	1.465	&	0.020 & 0.028 & 0.029 & 0.027	&	3 \hspace{0.2cm}\fcolorbox{black}{red}{\rule{0pt}{3pt}\rule{3pt}{0pt}} & \\
HD 199221                &	SAO 089278	&	G2V	&	20:55:02.41	&	+28 05 25.8	&	7.81	&	1.156	&	1.437	&	1.514	&	0.020 & 0.033 & 0.030 & 0.027	&	1 \hspace{0.2cm}\fcolorbox{black}{green}{\rule{0pt}{3pt}\rule{3pt}{0pt}} & \\
HD 353253                &	SAO 106663	&	G0	&	20:55:30.44	&	+19 41 11.0	&	9.21	&	1.099	&	1.433	&	1.483	&	0.020 & 0.044 & 0.050 & 0.028	&	1 \hspace{0.2cm}\fcolorbox{black}{green}{\rule{0pt}{3pt}\rule{3pt}{0pt}} & \\
HD 199898                &	SAO 164058	&	G2V	&	21:00:27.76	&	\textminus16 22 08.1	&	9.94	&	1.289	&	1.538	&	1.643	&	0.040 & 0.048 & 0.070 & 0.045	&	3 \hspace{0.2cm}\fcolorbox{black}{red}{\rule{0pt}{3pt}\rule{3pt}{0pt}} & \\
HD 200633                &	SAO 145075	&	G5V	&	21:04:44.15	&	\textminus04 49 44.0	&	8.34	&	1.140	&	1.427	&	1.509	&	0.010 & 0.021 & 0.033 & 0.022	&	1 \hspace{0.2cm}\fcolorbox{black}{green}{\rule{0pt}{3pt}\rule{3pt}{0pt}} & \\
HD 203311                &	SAO 164338	&	G2V	&	21:21:51.08	&	\textminus16 16 25.9	&	7.45	&	1.119	&	1.396	&	1.479	&	0.010 & 0.021 & 0.039 & 0.023	&	1 \hspace{0.2cm}\fcolorbox{black}{green}{\rule{0pt}{3pt}\rule{3pt}{0pt}} & \\
BD+22 4443               &	SAO 089861	&	G0	&	21:38:01.02	&	+22 49 08.5	&	9.32	&	1.157	&	1.389	&	1.434	&	0.020 & 0.028 & 0.039 & 0.028	&	1 \hspace{0.2cm}\fcolorbox{black}{green}{\rule{0pt}{3pt}\rule{3pt}{0pt}} & \\
BD+03 4598               &	SAO 126983	&	G0	&	21:40:41.80	&	+04 15 08.5	&	9.71	&	1.154	&	1.498	&	1.555	&	0.030 & 0.036 & 0.047 & 0.036	&	1 \hspace{0.2cm}\fcolorbox{black}{green}{\rule{0pt}{3pt}\rule{3pt}{0pt}} & \\
BD\textminus00 4251B              &	SAO 127005	&	F8	&	21:42:27.45	&	+00 26 20.3	&	9.14	&	1.189	&	1.472	&	1.559	&	0.020 & 0.028 & 0.047 & 0.033	&	1 \hspace{0.2cm}\fcolorbox{black}{green}{\rule{0pt}{3pt}\rule{3pt}{0pt}} & \\
HD 209793                &	SAO 107657	&	G5	&	22:05:52.20	&	+12 32 47.5	&	8.66	&	1.123	&	1.390	&	1.460	&	0.010 & 0.023 & 0.028 & 0.021	&	1 \hspace{0.2cm}\fcolorbox{black}{green}{\rule{0pt}{3pt}\rule{3pt}{0pt}} & \\
HD 210388                &	SAO 072067	&	G0	&	22:09:22.50	&	+35 07 45.3	&	7.47	&	1.129	&	1.388	&	1.463	&	0.009 & 0.023 & 0.026 & 0.017	&	1 \hspace{0.2cm}\fcolorbox{black}{green}{\rule{0pt}{3pt}\rule{3pt}{0pt}} & \\
BD+26 4382               &	SAO 090339	&	G0	&	22:13:47.31	&	+27 07 19.9	&	9.16	&	1.179	&	1.462	&	1.513	&	0.020 & 0.033 & 0.027 & 0.028	&	1 \hspace{0.2cm}\fcolorbox{black}{green}{\rule{0pt}{3pt}\rule{3pt}{0pt}} & \\
HD 211320                &	SAO 034223	&	G0	&	22:14:47.86	&	+57 42 38.7	&	8.62	&	1.076	&	1.369	&	1.458	&	0.020 & 0.044 & 0.035 & 0.034	&	1 \hspace{0.2cm}\fcolorbox{black}{green}{\rule{0pt}{3pt}\rule{3pt}{0pt}} & \\
HD 211476                &	SAO 107794	&	G2V	&	22:17:15.14	&	+12 53 54.6	&	7.04	&	1.160	&	1.455	&	1.498	&	0.020 & 0.027 & 0.026 & 0.026	&	1 \hspace{0.2cm}\fcolorbox{black}{green}{\rule{0pt}{3pt}\rule{3pt}{0pt}} & \\
BD+06 4993               &	SAO 127435	&	G0	&	22:17:37.41	&	+07 13 44.0	&	9.46	&	1.117	&	1.433	&	1.462	&	0.030 & 0.040 & 0.045 & 0.045	&	1 \hspace{0.2cm}\fcolorbox{black}{green}{\rule{0pt}{3pt}\rule{3pt}{0pt}} & \\
HD 212029                &	SAO 051893	&	G0	&	22:20:23.87	&	+46 25 05.8	&	8.51	&	1.137	&	1.427	&	1.441	&	0.010 & 0.038 & 0.033 & 0.025	&	1 \hspace{0.2cm}\fcolorbox{black}{green}{\rule{0pt}{3pt}\rule{3pt}{0pt}} & \\
HD 212083                &	SAO 165023	&	G3V	&	22:21:59.97	&	\textminus19 26 07.5	&	7.87	&	1.118	&	1.387	&	1.473	&	0.010 & 0.021 & 0.023 & 0.023	&	1 \hspace{0.2cm}\fcolorbox{black}{green}{\rule{0pt}{3pt}\rule{3pt}{0pt}} & \\
HD 212680                &	SAO 090497	&	G0	&	22:25:41.78	&	+24 06 01.4	&	8.95	&	0.775	&	0.951	&	0.956	&	0.020 & 0.027 & 0.028 & 0.026	&	3 \hspace{0.2cm}\fcolorbox{black}{red}{\rule{0pt}{3pt}\rule{3pt}{0pt}} & \\
HD 212816                &	SAO 165090	&	G3/5V	&	22:27:14.89	&	\textminus12 28 33.1	&	9.48	&	1.094	&	1.336	&	1.422	&	0.016 & 0.024 & 0.041 & 0.026	&	1 \hspace{0.2cm}\fcolorbox{black}{green}{\rule{0pt}{3pt}\rule{3pt}{0pt}} & \\
HD 215428                &	SAO 108146	&	G0	&	22:44:53.41	&	+18 01 43.8	&	8.50	&	1.122	&	1.336	&	1.410	&	0.010 & 0.034 & 0.022 & 0.022	&	1 \hspace{0.2cm}\fcolorbox{black}{green}{\rule{0pt}{3pt}\rule{3pt}{0pt}} & \\
HD 216505                &	SAO 165360	&	F7V	&	22:53:38.72	&	\textminus16 14 28.9	&	8.94	&	0.963	&	1.204	&	1.237	&	0.020 & 0.028 & 0.039 & 0.031	&	1 \hspace{0.2cm}\fcolorbox{black}{green}{\rule{0pt}{3pt}\rule{3pt}{0pt}} & \\
HD 217014                 &	SAO 090896	&	G2IV	&	22:57:27.98	&	+20 46 07.8	&	5.46	&	1.090	&	1.230	&	1.470	&	0.050 & 0.054 & 0.054 & 0.054	&	1 \hspace{0.2cm}\fcolorbox{black}{green}{\rule{0pt}{3pt}\rule{3pt}{0pt}} & \footnote{V\textminus J, V\textminus K colors from Vizier catalog II/225.}\\
HD 217443                &	SAO 127890	&	G0	&	23:00:37.13	&	+08 45 01.6	&	8.08	&	1.149	&	1.429	&	1.533	&	0.010 & 0.026 & 0.039 & 0.028	&	1 \hspace{0.2cm}\fcolorbox{black}{green}{\rule{0pt}{3pt}\rule{3pt}{0pt}} & \\
HD 217458                &	SAO 127897	&	F8/G0V	&	23:00:46.03	&	+03 20 37.3	&	8.59	&	1.103	&	1.385	&	1.462	&	0.020 & 0.031 & 0.037 & 0.034	&	1 \hspace{0.2cm}\fcolorbox{black}{green}{\rule{0pt}{3pt}\rule{3pt}{0pt}} & \\
HD 218647                &	SAO 146539	&	G1/2V	&	23:09:48.98	&	\textminus07 05 24.3	&	8.65	&	1.138	&	1.436	&	1.478	&	0.020 & 0.030 & 0.027 & 0.029	&	2 \hspace{0.2cm}\fcolorbox{black}{yellow}{\rule{0pt}{3pt}\rule{3pt}{0pt}} & \\
HD 220284                &	SAO 052943	&	G5	&	23:22:08.80	&	+49 32 01.6	&	7.90	&	1.184	&	1.443	&	1.517	&	0.010 & 0.022 & 0.021 & 0.019	&	1 \hspace{0.2cm}\fcolorbox{black}{green}{\rule{0pt}{3pt}\rule{3pt}{0pt}} & \\
HD 220500                &	SAO 073225	&	G0	&	23:23:54.23	&	+37 24 20.7	&	8.48	&	1.138	&	1.370	&	1.472	&	0.010 & 0.029 & 0.050 & 0.019	&	1 \hspace{0.2cm}\fcolorbox{black}{green}{\rule{0pt}{3pt}\rule{3pt}{0pt}} & \\
HD 220773                &	SAO 128181	&	G0	&	23:26:27.44	&	+08 38 37.8	&	7.10	&	1.099	&	1.388	&	1.449	&	0.010 & 0.022 & 0.034 & 0.020	&	1 \hspace{0.2cm}\fcolorbox{black}{green}{\rule{0pt}{3pt}\rule{3pt}{0pt}} & \\
HD 220845                &	SAO 073257	&	G5	&	23:26:49.69	&	+36 06 13.7	&	8.41	&	1.186	&	1.451	&	1.526	&	0.010 & 0.025 & 0.028 & 0.021	&	3 \hspace{0.2cm}\fcolorbox{black}{red}{\rule{0pt}{3pt}\rule{3pt}{0pt}} & \\
SA 115\textminus271               &	\textminus	&	F8	&	23:42:41.82	&	+00 45 13.1	&	9.69	&	1.176	&	1.434	&	1.555	&	0.001 & 0.020 & 0.044 & 0.021	&	1 \hspace{0.2cm}\fcolorbox{black}{green}{\rule{0pt}{3pt}\rule{3pt}{0pt}} & \\
HD 222788                &	SAO 108793	&	F3V	&	23:43:34.70	&	+19 07 47.4	&	9.08	&	0.809	&	0.985	&	1.029	&	0.020 & 0.031 & 0.034 & 0.031	&	3 \hspace{0.2cm}\fcolorbox{black}{red}{\rule{0pt}{3pt}\rule{3pt}{0pt}} & \\
HD 222814                &	SAO 146870	&	G2V	&	23:43:50.66	&	\textminus06 16 24.8	&	8.52	&	1.158	&	1.402	&	1.496	&	0.010 & 0.025 & 0.033 & 0.025	&	1 \hspace{0.2cm}\fcolorbox{black}{green}{\rule{0pt}{3pt}\rule{3pt}{0pt}} & \\
HD 223238                &	SAO 128385	&	G5V	&	23:47:52.41	&	+04 10 31.7	&	7.69	&	1.125	&	1.392	&	1.487	&	0.020 & 0.048 & 0.066 & 0.028	&	1 \hspace{0.2cm}\fcolorbox{black}{green}{\rule{0pt}{3pt}\rule{3pt}{0pt}} & \\
BD+17 4993               &	SAO 108869	&	G0	&	23:51:14.65	&	+18 21 32.3	&	9.41	&	1.115	&	1.404	&	1.438	&	0.020 & 0.028 & 0.030 & 0.034	&	1 \hspace{0.2cm}\fcolorbox{black}{green}{\rule{0pt}{3pt}\rule{3pt}{0pt}} & \\
HD 224465 &	SAO 035934	&	G4V	&	23:58:06.82	&	+50 26 51.6	&	6.64	&	1.148	&	1.407	&	1.499	&	0.020 & 0.028 & 0.060 & 0.027	&	1 \hspace{0.2cm}\fcolorbox{black}{green}{\rule{0pt}{3pt}\rule{3pt}{0pt}} &
\enddata
\end{deluxetable*}
\end{longrotatetable}

\end{document}